\documentclass[draft]{agujournal2019}
\usepackage{url} %this package should fix any errors with URLs in refs.
\usepackage{lineno}
\usepackage{soul}
\usepackage{amsmath}
\usepackage{amssymb}
\usepackage{amsfonts}
\usepackage{bm}
\usepackage{graphicx}
\usepackage{caption}

\draftfalse

\journalname{Journal of Advances in Modeling Earth Systems (JAMES)}

\begin{document}

\title{ACE2-NEMO: Coupling an ML atmospheric emulator to a full-depth dynamical ocean model}

\authors{Bobby Antonio\affil{1}, Kristian Strommen\affil{1,2}, Pablo Ortega\affil{3}, Hannah M. Christensen\affil{1}}

\affiliation{1}{Department of Physics, University of Oxford, Oxford, UK}
\affiliation{2}{European Centre for Medium-Range Weather Forecasts, Shinfield Rd, Reading, RG2 9AX, United Kingdom}
\affiliation{3}{Barcelona Supercomputing Center (BSC), Barcelona, Spain}

\correspondingauthor{Bobby Antonio}{bobby.antonio@physics.ox.ac.uk}

\begin{keypoints}
\item We introduce a novel hybrid modelling framework, by coupling a data-driven atmospheric emulator to a full-depth physical ocean model.
\item The model shows a realistic mean state, seasonal cycle, and temporal evolution, with no noticeable drift at a global scale.
\item Unusual variability and CO\textsubscript{2} response point to problems with atmospheric feedback and short-wave radiation in ACE2.
\end{keypoints}

\begin{abstract}
   Understanding how fast atmospheric variability shapes slow climate variability and sensitivity remains a central challenge in Earth-system science. Recent advances in machine-learned (ML) atmospheric models have demonstrated remarkable skill on weather timescales, but their emergent behaviour in a fully coupled climate system remains largely unexplored. We present early results from a new hybrid modelling framework, in which the ACE2 ML atmospheric emulator is interactively coupled to the NEMO ocean model. We report on a set of 70-year coupled simulations (1950–2020 historical forcing and fixed-1950s control). These experiments represent, to our knowledge, one of the first multi-decadal integrations of a machine-learned atmosphere interacting with a full-depth dynamical ocean. Several historical and fixed-1950s control simulations from the fully dynamic global coupled climate model EC-Earth, which has the same ocean component used in ACE2-NEMO, are also considered for comparison. We assess the behaviour of the coupled system, with particular focus on low-frequency tropical variability and the climate response to greenhouse-gas forcing. Analysis of potentially emergent El Ni\~{n}o-like variability reveals realistic fast timescale air-sea coupling in the tropical Pacific, but the temporal variability is unrealistic, with very low amplitude oscillations; this appears to be due to weak atmospheric feedback in the tropical Pacific. The response to CO\textsubscript{2} forcing shows initial agreement with EC-Earth3P but deviates due to reduced downward short-wave radiation in ACE2. These results provide a unique test of physical realism and generalisability for atmospheric emulators and evaluate the possible role of entirely machine-learned components in next-generation Earth system models.
\end{abstract}

\section*{Plain Language Summary}

To model the Earth's climate, it is important to create computer models that accurately capture the complex behaviour of the atmosphere and ocean. Data-driven models learn to emulate the Earth's weather and climate from large amounts of data and have recently been shown to be effective at producing more accurate forecasts compared to conventional models that use the laws of physics. Most of these data-driven models focus on the weather and only model the atmosphere, not including interactions with the ocean, which are important to model the climate. We develop a new type of model that connects an existing data-driven model of the atmosphere to a conventional ocean model and explore how realistic this model is. We find that, whilst the model is stable and realistic in many ways, the model cannot capture important phenomena, such as the response to CO\textsubscript{2} in the atmosphere. Overall, this demonstrates a new way of modelling the Earth's climate, and highlights problems with existing data-driven models.

\section{Introduction}

A central challenge of environmental science is to understand how processes acting at different timescales interact to produce emergent climate variability and sensitivity. Hasselmann's key insight was that rapid atmospheric fluctuations are integrated by slower earth system components, such as the ocean and cryosphere, to produce long-term climate variability \cite{hasselmann_stochastic_1976}, whilst Manabe showed how fast water vapour and cloud feedbacks determine the sensitivity of the system to increased CO\textsubscript{2} forcing \cite{manabe1967thermal}.

General circulation models (GCMs) have provided a core method to explore these multi-scale interactions. Despite significant advances in accurately modelling the earth system, notable biases still remain in the representation of atmospheric features such as clouds, convection, and precipitation \cite{li2014tropical, hwang2013link, hyder2018critical}. Improved parameterizations, including using machine learning, provide one way to improve this \cite{gentine_deep_2021, grundner_deep_2022, behrens_simulating_2025, bretherton_correcting_2022, miller2025characterizing}, although problems remain in achieving online stability \cite{brenowitz2020machine}. Higher resolution models could also reduce these biases \cite{palmer2014climate}, but require substantially increased computational resources, which may limit our ability to incorporate additional complexity or explore many diverse emissions pathways. 

Recently, a new paradigm of atmospheric modelling has emerged through machine learning (ML), where complex statistical relationships between variables are learned in place of explicitly modelling physical processes and parameterizing subgrid processes. ML atmospheric emulators trained on reanalysis and observations have demonstrated an impressive ability to model the atmosphere at short timescales, with most applications focusing on medium-range \cite{lam_learning_2023, bi_accurate_2023, price_probabilistic_2025, kurth2023fourcastnet} out to seasonal timescales \cite{kent_skilful_2025, antonio_seasonal_2026, zhang_advancing_2025}, with models achieving state-of-the art without any sea surface inputs, or using prescribed sea surface temperatures.

The success of ML models at forecasting the weather, and recent demonstrations of atmospheric emulators stable at climate timescales \cite{brenowitz_climate_2025, watt-meyer_ace2_2025, chapman_camulator_2025, kochkov_neural_2024, cachay_probabilistic_2024, brenowitz_climate_2025, guan_lucie_2025}, motivates the pursuit of using ML in climate simulations, where they have the potential to provide increased speed, larger ensembles, and lower biases \cite{eyring_pushing_2024, bracco_machine_2025}. To realise this potential, we must move towards ML emulations that couple different parts of the earth system and deal with interactions across timescales. Coupling ML atmospheric emulators to ocean models over multi-decadal simulations also provides a test of the physical realism of the models, and reveals to what extent they have learned the complex atmospheric feedbacks that lead to realistic climate sensitivity and coupled modes of variability such as the El Ni\~{n}o Southern Oscillation. 

To date, there have been relatively few demonstrations of machine learning models emulating coupled atmosphere-ocean dynamics at multi-decadal timescales with a deep ocean; in \citeA{duncan_samudrace_2025} this was successfully achieved by training atmospheric and oceanic ML emulators separately, before fine-tuning them together. Training ML ocean emulators is potentially more challenging compared to their atmospheric counterparts and remains an active area of research \cite{dheeshjith_samudra_2025, el_glonet_2025, subel_building_2024, wang_xihe_2024, cui_forecasting_2025}. One reason is a sparsity of observations, which, combined with the relatively slow evolution of the ocean, makes it challenging for an emulator trained on observational datasets to faithfully represent ocean variability, particularly in the deep ocean. For example, high quality observations for the Atlantic meridional overturning circulation, a key ocean phenomenon varying on multi-decadal to centennial timescales, are only available from 2004 onwards \cite{moat2026atlantic, fu2025characterizing}. In addition, the ocean presents complex fluid dynamics due to differences in density and salinity, and added complexities from coastal and sea floor boundaries. We note that the coupled ML emulators in \citeA{duncan_samudrace_2025} also use multi-day ocean timesteps, which may limit their ability to capture phenomenon that are affected by sea surface diurnal temperature, such as the Madden-Julian oscillation \cite{kawai_diurnal_2007, seo_coupled_2014}.

Motivated by the above, we present the first coupling of a machine-learnt emulator of the atmosphere (ACE2 \cite{watt-meyer_ace2_2025}) to a full-depth dynamical ocean model (NEMO \cite{madec_nemo_2023}), without any fine-tuning or retraining of the atmospheric model. This provides a unique test of how well the atmospheric emulator can respond to ocean data that is very different to its training data, provides a stringent test of how well the atmosphere has learned the appropriate feedbacks to represent coupled variability and forced response, and tests how well it can be used as a `general purpose' component of a climate model. Coupling these two very different models poses several technical challenges; NEMO runs on multiple CPUs, whilst ACE2 runs on a GPU, and conventional HPC environments do not easily facilitate communication between the two. ACE2 does not predict certain fluxes that need to be estimated. We also find that modifications of fluxes over sea ice are required, to avoid large drifts in sea ice volume and sea surface height.

The coupled model runs stably for at least one hundred years, and our results show that the model produces a realistic mean state with no obvious signs of instability despite the out-of-sample ocean inputs, and differences that are most likely due to different initial conditions and modifications to the sea ice coupling. To investigate the model's ability to simulate coupled variability, we explore how well ACE2-NEMO represents El Ni\~{n}o Southern Oscillation (ENSO). Our analysis reveals that, whilst ACE2-NEMO produces realistic spatial correlations of precipitation, sea level pressure and winds, the Ni\~{n}o 3.4 index time series of ACE2-NEMO is extremely muted compared to observations and lacks realistic temporal variability. Further investigation reveals significant differences in the response of surface winds and radiative cooling to sea surface temperatures, which we hypothesise cause differences in feedback that lead to reduced growth and persistence of El Ni\~{n}o/La Ni\~{n}a states.

Historical simulations also reveal that, whilst global 2-metre temperature evolves realistically until 1980, the sea surface temperature shows an unusual decrease beyond this; this appears to be due to an overestimation in ACE2 of the decrease in downward short-wave radiation.

This work represents a first step in a novel hybrid approach to incorporating machine learning into climate modelling. Apart from providing a deeper insight to the reliability and realism of machine-learnt atmospheric emulators, this approach has the potential to be applied in other problems in climate modelling, such as faster coupled spin-up of the ocean, and coupled seasonal forecasts using machine learning.

\section{Methods}

\begin{figure}[!t]
    \includegraphics[width=\textwidth]{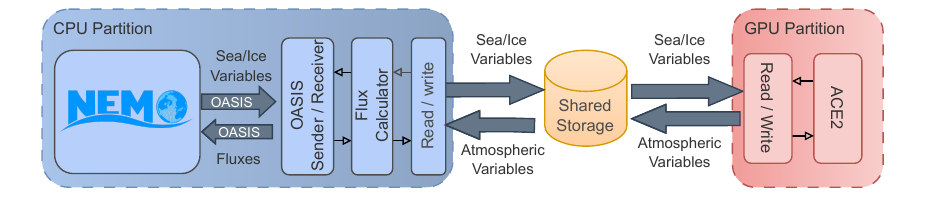}
       \caption{Schematic diagram of the infrastructure created to couple NEMO to ACE2. }
    \label{fig:coupling_diagram}
\end{figure}

\subsection{The ACE2-NEMO model}
\subsubsection{Atmospheric model}

We use ACE2 as the atmospheric model \cite{watt-meyer_ace2_2025}. ACE2 is a machine-learnt climate emulator that has been demonstrated to be stable out to thousands of model years. The model checkpoint we use is trained on ERA5 data \cite{ai2_2025,hersbach_era5_2020}, forced by global mean atmospheric CO\textsubscript{2}, top of atmosphere short-wave radiation flux, and sea surface temperature, and without any aerosol forcing or time information (e.g.~year, day of year). The input and output of the model is on a regular $1^{\text{o}}\times1^{\text{o}}$ grid at 8 different levels, with predictions made autoregressively in 6-hour timesteps. ACE2 predicts many atmospheric variables, including sensible, latent, short-wave and long-wave surface fluxes; these fluxes are predicted as averages over 6 hours.

In order to pass NEMO outputs to ACE2, changes to ACE2 are required. For ingestion of sea surface temperature ($T_{\text{oce}}$) and sea ice temperature ($T_{\text{ice}}$), we introduced a modified ocean class in ACE2, to allow reading ocean information from a file using a polling mechanism. Once the data was read, the surface temperature $T_{\text{surf, oce}}$ over ocean points was overwritten using a weighted sum of $T_{\text{oce}}$ and $T_{\text{ice}}$, with weighting given by the sea ice fraction $f_{\text{ice}}$ output from NEMO:
\begin{align}
T_{\text{surf, oce}} = f_{\text{ice}}T_{\text{ice}} + (1-f_{\text{ice}})T_{\text{oce}}
\end{align}
Passing the sea ice fraction from NEMO to ACE2 required a different mechanism, since sea ice fraction is an input-only forcing field in ACE2, which cannot be modified via the ocean class. We instead made use of the `perturbation` mechanism created for e.g. perturbing the sea surface temperatures; by constructing a custom perturbation function in ACE2 that reads data from disk, we are able to overwrite the sea ice fraction forcing data sent to ACE2 from the forcing files with the data from NEMO.

% In order to provide ACE2 with outputs from the ocean model, we utilise the infrastructure already in place for coupling to a slab ocean, and for perturbing the forcing inputs. The fields from the ocean are first written to shared storage (see Sec. \ref{sec:coupling}
% ); these are picked up by ACE2 and used to overwrite the surface temperature and sea ice fraction fields (see suppleme

\subsubsection{Ocean model}
We use the Nucleus for European Modelling of the Ocean (NEMO) model \cite{madec_nemo_2023}, which includes a sea ice model. The ocean grid is chosen as ORCA1, in order to closely match the $1^{\text{o}}$ grid of the atmosphere, with 75 depth levels, and a model time step of 45min. 

To align the model as closely as possible with the EC-Earth simulations used for comparison and ocean initial conditions, we use NEMO version 3.6. Configuration files for NEMO are taken from those used within the PRIMAVERA experiments \cite{haarsma_highresmip_2020} to align with existing control and historical runs starting in 1950. For run-off and geothermal heating inputs to NEMO, we use the standard climatological data used within EC-Earth3. NEMO is initialised from the 1951 restart from the historical run performed in \citeA{bilbao_assessment_2020}, which starts in 1850 (see Sec.~\ref{sec:data}).

To facilitate robust compilation and setup of the model, we use the infrastructure created for EC-Earth4 (\url{https://ec-earth-4-docs.readthedocs.io/}). This utilises python packages such as scriptengine (\url{https://scriptengine.readthedocs.io}) and rdy2cpl (\url{https://github.com/uwefladrich/rdy2cpl}) to reliably set up and compile the files required for NEMO and OASIS to run.

\subsubsection{Coupling infrastructure}
\label{sec:coupling_infra}

A fundamental challenge is that NEMO runs on multiple CPUs, whilst ACE2 runs on a single GPU, and the conventional HPC environments do not allow multinode hybrid CPU-GPU jobs within the same MPI environment, thus ruling out a straightforward heterogeneous batch job. As a workaround we used shared storage to mediate data passing between ACE2 and NEMO (see Fig.~\ref{fig:coupling_diagram}), whereby NetCDF files are written to a temporary folder during the run, and a polling mechanism is used to check for the presence of the correct file. This adds some read/write overhead; however, this was small compared to the overhead of running the models and calculating the fluxes. 

To communicate with NEMO, we use the existing integration of NEMO with the OASIS library \cite{craig_development_2017}. Apart from allowing sharing of variables, OASIS also provides the regridding utilities used to transform between the NEMO ORCA1 grid and the regular ACE2 grid, with setup facilitated by the \verb|rdy2cpl| package (see Table S1 for details of regridding methods used). Using PyOASIS \cite{gambron_pyoasis_2021}, SST, bulk ice temperature, sea ice fraction, sea ice thickness, ocean currents, and ice velocities from NEMO are intercepted by a Python router module, within which atmospheric variables are also read from shared storage. The coupling frequency is set to the timestep of ACE2, 6 hours (see Fig.~S1 for a detailed diagram of how data is exchanged during coupling).

The freshwater fluxes that NEMO receives are evaporation (over ocean and ice points) and precipitation (liquid and solid). The heat fluxes NEMO receives are non-solar heat flux (sum of sensible, latent, and long-wave fluxes), solar heat flux, and momentum fluxes, all provided for ocean and sea ice separately. The version of ACE2 available to us at the time of writing \cite{ai2_2025} produces most of the fluxes, except for momentum fluxes, evaporation, and solid precipitation. To calculate momentum fluxes and evaporation, we use the AirSeaFluxCode package \cite{biri_airseafluxcode_2023}, a python implementation of the bulk formulae used within NEMO for calculating fluxes when forced by atmospheric variables \cite{brodeau_climatologically_2017}. This parameterises the turbulent heat, momentum, and moisture exchanges at the air-sea boundary. Fluxes are expressed in terms of transfer coefficients estimated iteratively using Monin-Obukhov Similarity Theory, that allow conversion of surface gradients into fluxes. There are a diverse range of formulations of these transfer coefficients: we use the ECMWF scheme to make the momentum flux calculations as consistent as possible with ECMWF's ERA5 data that ACE2 is trained on. The output of AirSeaFluxCode also provides latent heat of vaporisation, which allows conversion of latent heat flux to evaporation. As input, the flux calculation takes in surface state variables: wind speed magnitude (relative to surface currents), 2m temperature, surface humidity, mean sea level pressure, sea surface temperature, downward short- and long-wave radiation. Since sea surface temperature from NEMO is bulk temperature, but the ECMWF scheme requires skin temperature, a cool-skin/warm layer adjustment is calculated as part of the routine. The maximum number of iterations is set to 50. 

Since the bulk transfer coefficients are calculated using the atmospheric conditions of ACE2, including ACE2 surface winds, using a scheme that is similar to that used to create the ERA5 dataset, this calculation ensures the momentum fluxes are as consistent as possible with ACE2’s atmospheric state. A verification of the flux calculation is provided in Sec.~S10, using ERA5 and ACE2 as inputs: this shows generally low biases on the fluxes calculated using AirSeaFluxCode. The largest biases are seen for evaporation, which shows increased evaporation relative to ERA5 when using ERA5 as input, and decreased evaporation relative to ERA5 when using ACE2 variables as input. 

To provide NEMO with solid precipitation, we implemented a simple heuristic model to estimate it. Based on observations of the relationship between solid precipitation and surface temperature with ERA5, we set the solid precipitation to be equal to the total precipitation over sea points whenever the 2-metre temperature is $\leq 273\text{K}$. Whilst this obtains solid precipitation that is reasonably consistent with ERA5, there are notable biases (see Sec.~S10), so that improving upon this heuristic is a focus of future model development. River discharge is provided by standard climatology files from within the EC-Earth3 data. 

Early testing of NEMO coupled to ACE2 showed large unrealistic downward trends in global sea surface height, and large upwards trends in global sea ice volume (Fig.~S6). We observed that this decrease was partly removed when freshwater fluxes were removed and therefore applied the setting in NEMO that enforces freshwater flux conservation (\verb+nn_fwb=1+ in the NEMO ocean namelist), so that evaporation minus the sum of precipitation and run-off is adjusted to be zero at each time step. Although ACE2 has freshwater fluxes conserved at each time step, such a freshwater imbalance is to be expected, because the model uses run-off climatology rather than a fully coupled run-off model. This adjustment is applied within NEMO by calculating the global freshwater imbalance and distributing this across all points weighted by grid cell area. The magnitude of this correction is quantified by comparing ACE2-NEMO-control to a short model run that has freshwater conservation turned off; this reveals a sea surface height drift of around -1m/year without this correction.

\subsubsection{Sea ice coupling}
\label{sec:sm_shf}

Despite imposing a freshwater conservation constraint (see Sec.~\ref{sec:coupling_infra}), the model still showed substantial drift in sea surface height and sea ice volume. We observed that over many ice areas, the direction of the sensible heat flux did not align with the gradient between 2-metre temperature and sea ice temperature (Fig.~S12). This problem persisted when running ACE2 with prescribed sea surface variables more aligned with 1950s ERA5 conditions (see Sec.~S6 in supplementary material), suggesting this is due to out-of-sample inputs from NEMO to ACE2, or because NEMO outputs bulk ice temperature rather than skin temperature over ice. A solution which removed the drift and aligned the flux direction with the temperature gradient was to calculate the fluxes directly using the simple CORE bulk formulae over ice \cite{large_diurnal_2004, large_global_2009}, as implemented in the NEMO 3.6 code (within the \verb+sbcblk_core+ module).

The bulk formulae to calculate momentum flux, sensible heat flux, and latent heat flux over ice, denoted $\boldsymbol{\tau}_{\text{ice}}$, $H_{\text{ice}}$ and $L_{\text{ice}}$ respectively, are:
\begin{align}
\boldsymbol{\tau}_{\text{ice}} &= \rho C_{\text{ice}} (\mathbf{u}_{10m}-\mathbf{u}_{\text{ice}}) |\mathbf{u}_{10m} - \mathbf{u}_{\text{ice}}|\\
H_{\text{ice}} &= \rho C_{\text{ice}} c_p (T_{2m} - T_{\text{ice}} ) |\mathbf{u}_{10m} - \mathbf{u}_{\text{ice}}|\\
L_{\text{ice}} &= L_s \rho C_{\text{ice}} (q_{2m} - 11637800  \exp( -5897.8 / T_{\text{ice}} ) / \rho) |\mathbf{u}_{10m} - \mathbf{u}_{\text{ice}}|
\end{align}
Where $\rho=1.22$ kg m\textsuperscript{-3} is the air density, $c_p=1005.0$ J kg\textsuperscript{-1} K\textsuperscript{-1} the specific heat capacity of moist air, $L_s=2.839 \times 10^6$ J kg\textsuperscript{-1} the latent heat of sublimation, and $C_{ice}=1.4 \times 10^{-3}$ is the heat transfer coefficient. $T_{2m}$, $\mathbf{u}_{10m}$ and $q_{2m}$ are atmospheric fields, namely 2-metre temperature, 10-metre wind velocity, and 2-metre specific humidity respectively, predicted by ACE2. $T_{\text{ice}}$ is the bulk temperature of the ice, and $\mathbf{u}_{\text{ice}}$ is the velocity of the ice, both output from NEMO. Fluxes are defined such that a positive flux goes downwards (i.e.~from air to ocean) to be consistent with NEMO conventions. Latent heat fluxes are constrained to be negative, following the implementation of the CORE bulk formulae used in NEMO in \verb|sbc_blkcore|.

Net long-wave and short-wave radiation, $R_{\text{lw},\text{ice}}$ and $R_{\text{sw},\text{ice}}$ respectively, are calculated as:
\begin{align}
   R_{\text{lw},\text{ice}} &= 0.95 ( R_{\text{lw} \downarrow} - \sigma_B T_{\text{ice}}^4 )\\
   R_{\text{sw},\text{ice}} &= (1-\alpha_{\text{ice}}) R_{\text{sw} \downarrow}
\end{align}
where $\sigma_B$ is the Stefan-Boltzmann constant, $R_{\text{lw} \downarrow}, R_{\text{sw} \downarrow}$ are the downwelling long- and short-wave radiative fluxes provided by ACE2, and $\alpha_{\text{ice}}$ is the ice albedo provided by NEMO.

Large values of sea ice thickness near coastal points were also observed, causing the model to crash due to large sea surface height values. These appeared to be due to large fluxes produced because of the temperature in the coastal waters, probably resulting from the regridding of the fluxes to the ORCA1 grid. To fix this, we removed heat fluxes for any point within 1 grid square of the coast that also had more than 10\% coverage of sea ice. This affects around 3\% of the sea ice by area, where grid cells are deemed to be affected if the fluxes are masked at that location more than 10\% of the time (see also Fig.~S16). The effect on the polar heat budgets was estimated by comparing to a 1-year long run that used the unmasked ACE2-NEMO ice fluxes; compared to this run, ACE2-NEMO-control reduced both solar and non-solar heat fluxes over ice by around -12\%: however, we expect this to be a conservative estimate, since the unmasked fluxes contain large fluxes that may skew the results. This modification of the polar heat fluxes may be a contributing factor to the large annual variability in sea ice volume (see Sec.~S8).

\subsubsection{Computational Cost}
\label{sec:compute_cost}

ACE2-NEMO runs at a rate of around 10 simulated years per day (SYPD), compared with around 12 SYPD for NEMO alone using the same resources. With this setup, NEMO typically runs at around 20s per simulated day, whilst ACE2 takes less than 1s per simulated day. The interface between ACE2 and NEMO is dominated by the speed of flux calculations using AirSeaFluxCode (20-40s per simulated day) which is not currently optimised for this application; note that if ACE2 provided momentum fluxes and an estimate of latent heat of vaporization, this cost would be removed.

The NEMO model is run in parallel across 2 CPU nodes, with each node running 16 tasks. The Python router runs on a separate node, although this router is lightweight and so has negligible cost compared to NEMO. ACE2 is run on a single NVIDIA Ampere A100 GPU. Following a similar calculation to \citeA{pathak_fourcastnet_2022}, the NEMO model uses approximately as many cores as are available on one Cray XC40 node and so takes 20 node-seconds per simulated day, equating to around 5.4kJ per simulated day. ACE2 takes 1 node-second per simulated day, equating to around 0.4kJ, based on a thermal design power of 300W \cite{nvidia_product_brief}; therefore ACE2-NEMO consumes around 5.8kJ per simulated day. EC-Earth3 requires around three times as many node-seconds per simulated day as NEMO by itself \cite{acosta_balancing_2023}, so that EC-Earth3 consumes around 16.2kJ per simulated day. ACE2-NEMO has the potential to be significantly more efficient to run than EC-Earth3, requiring around a third of the energy per simulated day, with further efficiency gains expected after optimisation of the flux calculation. We note however that this simple energy proxy based on node-seconds and thermal design power is relatively crude and does not account for disk read/write operations; we plan to perform more stringent efficiency and scaling experiments in future work. 

\subsection{Benchmark Models}
\label{sec:data}

Two versions of EC-Earth are used for comparison with ACE2-NEMO for two main reasons: it also uses NEMO as its ocean component, and its atmospheric model is the Integrated Forecast System (IFS) from the European Centre for Medium-Range Weather Forecasts (ECMWF), the atmospheric model used to to construct the ERA5 dataset that ACE2 is trained on. Several different EC-Earth3 simulations are used to compare with; we use 1950s control and hist-1950 runs performed following the HighResMIP protocol with the PRIMAVERA version of EC-Earth \cite{haarsma_highresmip_2020}, denoted ECE3P-control and ECE3P-hist respectively. These have the same ocean configuration as ACE2-NEMO, but start from different ocean initial conditions, which come from a previous 50-year spin-up. We also compare ACE2-NEMO simulations with a historical run starting from 1850 with the CMIP6 version of EC-Earth \cite{bilbao_assessment_2020, doscher_ec-earth3_2022} denoted ECE3-historical, which also provides the ocean initial condition for our experiments. Using initial conditions from the 1850s historical run means that we are likely to have a well-spun-up ocean, reducing ocean drifts in NEMO due to residual spin-up, compared to using the 50-year spun-up initial conditions from the PRIMAVERA experiment. Since the initial conditions are from an ocean model with a different configuration, we also expect this to lead to some initial adjustments in NEMO during the simulation. All of the EC-Earth3 simulations use the T255 atmospheric grid ($\sim80\text{km}$) and the ORCA1 ocean grid ($\sim100\text{km}$). ACE2 atmospheric initial conditions are taken from the pre-prepared initial conditions detailed in \citeA{watt-meyer_ace2_2025}. 

To compare our simulations with SamudrACE, we perform a 70-year simulation using the SamudrACE-CM4-piControl model, trained on  the Geophysical Fluid Dynamics Laboratory's Climate Model v4 \cite{duncan_samudrace_2025}. At inference time, we use initial conditions from the first year of the held-out test period, with constant forcing taking from the same period.

We include results from an AMIP-style run, where ACE2 uses prescribed sea and ice fields taken from the ECE3P-control run (labelled ACE2-ECE3P-prescribed). This uses hourly data for the sea surface temperature, and daily data for sea ice concentration and temperature. Since the ECE3P-control run is initialised much closer to ERA5 in the 1950s, this allows an assessment of how well ACE2 may perform without the warm ocean conditions, and without any biases introduced by the coupling calculation.

\section{Results}

Two main sets of experiments were performed using ACE2 coupled to NEMO on an ORCA1 grid; a set of control runs using constant 1951 CO\textsubscript{2} forcing (ACE2-NEMO-control), and a set of historical runs using historical CO\textsubscript{2} forcing from 1951 to 2021 (ACE2-NEMO-hist). In both cases a 3-member lagged ensemble was created by performing runs with atmospheric initial conditions shifted by 1 day. These are compared with three existing EC-Earth3 simulations; a historical run starting from 1850s from \citeA{bilbao_assessment_2020}, denoted ECE3-historical, that provides the 1950 initial conditions used in the ACE2-NEMO simulation, and 1950s control and historical runs with the PRIMAVERA version of EC-Earth \cite{haarsma_highresmip_2020}, denoted ECE3P-control and ECE3P-hist respectively. Further details on the models, their implementation and the data used are given in Sec.~\ref{sec:data}.

\subsection{Model stability and mean state}

\begin{figure}[!t]
\centering
    \includegraphics[width=0.98\textwidth]{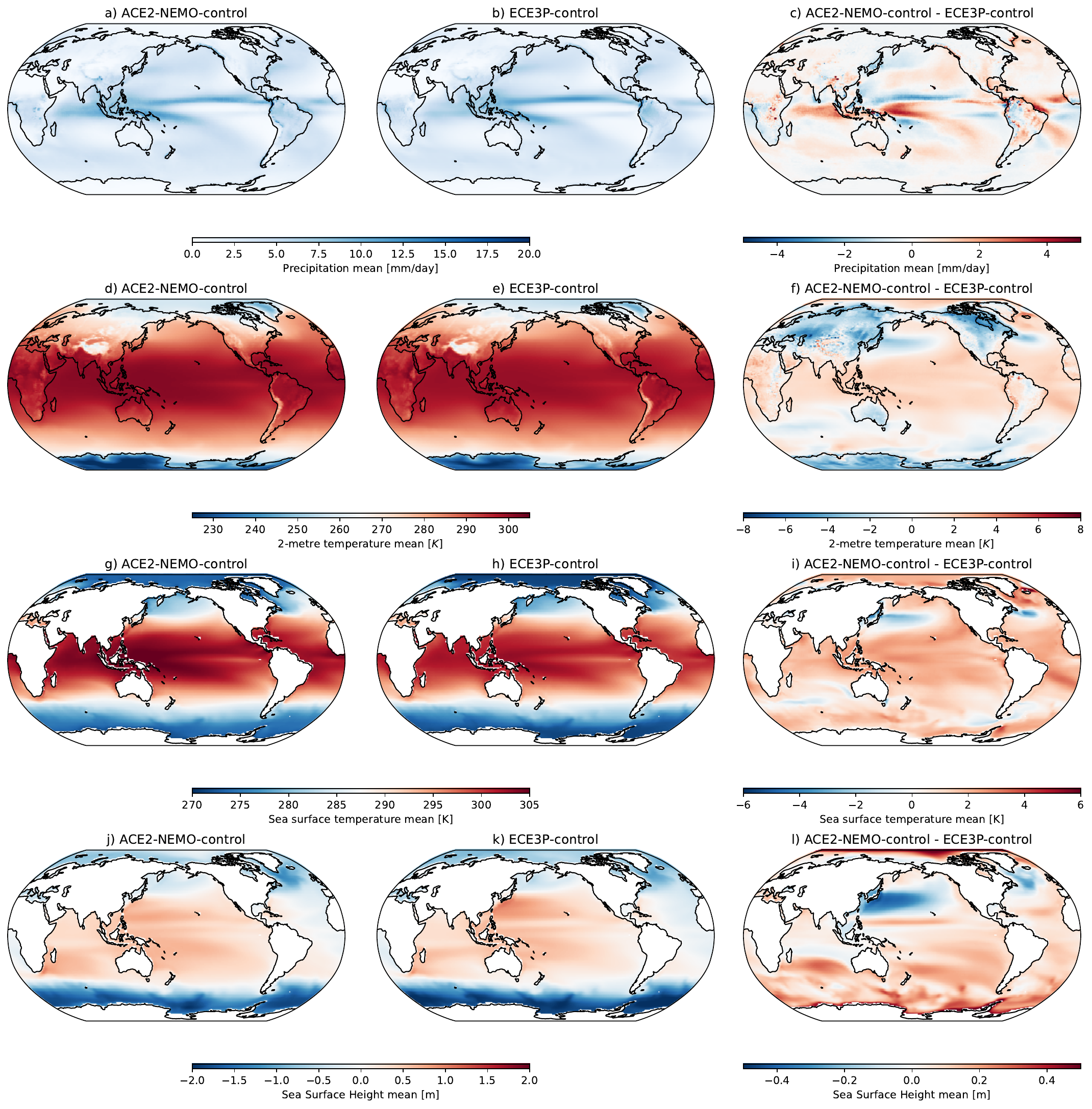}
       \caption{Time-averaged variables for the first ensemble member of the 70-year ACE2-NEMO-control and ECE3P-control experiments (a and b) daily precipitation (c and d) 2-metre temperature (e and f) sea surface temperature (g and h) and sea surface height. }
    \label{fig:mean_state}
\end{figure}

When coupling ACE2 to NEMO, we provide it with ocean inputs (in terms of boundary conditions) that are out of sample from what it has been trained on, which might induce some adjustments or drifts. Therefore, we begin by examining the stability of the coupled model and investigate the extent to which it has reasonable behaviour, or develops persistent long-term trends indicative of potential drifts. We compare ACE2-NEMO-control with ECE3P-control. These simulations are not expected to line up precisely, since the ECE3P-control run starts from different ocean initial conditions, and uses a dynamical atmosphere model that ACE2 has not been trained to emulate. However, we can still compare the mean state behaviours to look for differences in patterns or trends.

Plots of climatological precipitation, 2-metre temperature, sea surface temperature and sea surface height are shown in Fig.~\ref{fig:mean_state}. The spatial distribution of these variables is in good agreement with that seen for ECE3P-control. However, 2-metre temperature and SST are warmer for ACE2-NEMO-control (panels f and i); one reason for this is the ocean initial conditions used are around 0.5-1 K warmer than the ECE3P-control initial state (see Fig.~S2), and additionally there is warming seen during the first year that appears to be due to an imbalance in evaporation and the air-sea temperature gradient (see Sec.~S2). There are also differences in the sea surface height (panel l), particularly noticeable in the western Pacific; this may be related to differences in balance of freshwater fluxes, or thermal expansion. Differences in the northern Atlantic, Arctic Ocean, and coastal Antarctica may reflect the use of CORE bulk formulae over ice and the coastal ice flux masking. These mean state differences are further investigated by comparing ACE2-NEMO-control to ACE2-ECE3P-prescribed; the results (shown in Sec.~S6) confirm the biases in 2-metre temperature are significantly reduced and are of the opposite sign for the AMIP run. Heat and freshwater fluxes averaged over the control run show good visual agreement with ECE3P-control (Fig.~S3), confirming that ACE2 is still able to produce realistic fluxes despite being provided with NEMO ocean inputs.

The seasonal cycle of sea surface temperature and surface fluxes in different latitudinal bands also aligns well with ECE3P-control, but with some differences in the variance of seasonal cycle for the sea surface temperature in the extratropics, heat fluxes in the tropics, and momentum fluxes in the extratropics (Fig.~S4). A comparison of interannual variability shows broadly comparable ranges of variability, with ACE2-NEMO showing larger ranges of heat fluxes, and offsets in the median fluxes compared to ECE3P-control, particularly for the momentum fluxes (Fig.~S5).

\begin{figure}[!t]
\centering
    \includegraphics[width=\textwidth]{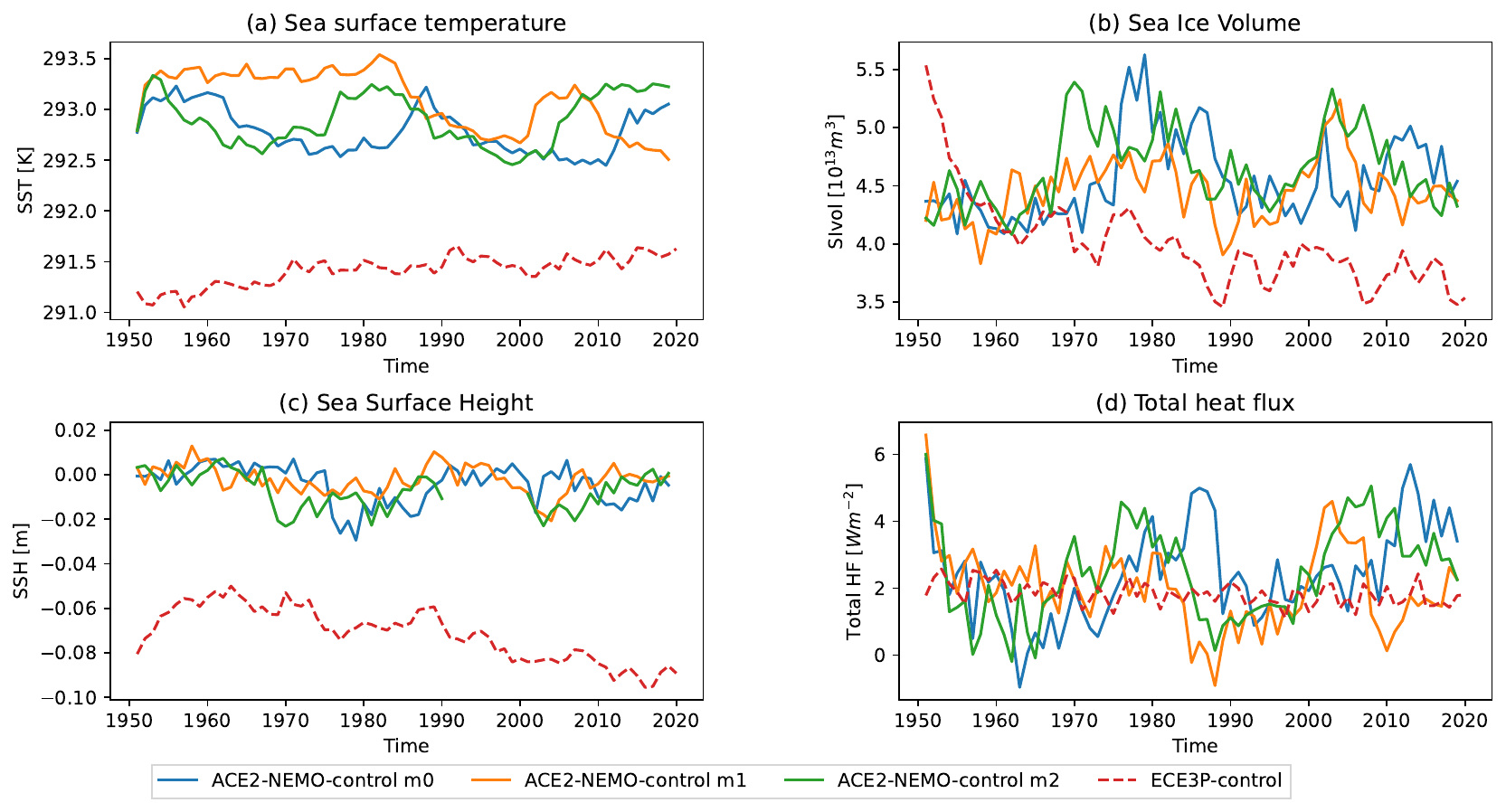}
       \caption{Globally averaged time series for the ACE2-NEMO-control 3-member ensemble and EC-Earth3P-control. Cosine-latitude weighting is used to account for different grid cell size (a) sea surface temperature, (b) sea ice volume, (c) sea surface height and (d) total heat flux (non-solar plus solar heat fluxes).}
    \label{fig:global_drift}
\end{figure}

Time series of global averages of several selected variables are shown for the control runs in Fig.~\ref{fig:global_drift}. The global sea surface temperatures (panel a) show little overall trend, although a larger multi-decadal variability than ECE3P-control is evident with a possible oscillation between two different modes, which can be seen in all three members. The ACE2-NEMO SSTs are also warmer than ECE3P-control by around 1-2 K, due to the difference in initial conditions and increased evaporation mentioned above. Sea ice volume (panel b) shows a much larger change for ECE3P-control than ACE2-NEMO, with an initial decrease in volume of around one third over 20 years, perhaps due to the shorter spin-up of ECE3P-control compared to the ECE3-historical initial conditions used for ACE2-NEMO. ACE2-NEMO shows a slight increase in volume over the same period, such that both models converge on approximately the same amount. A breakdown of sea ice volume by hemisphere is shown in Fig.~S12; this shows that the initial decrease in sea ice volume for ECE3P-control is from changes in the Arctic sea ice, whilst the slight increase for ACE2-NEMO-control is driven by changes in the Antarctic sea ice. Global sea surface height (panel c) shows a very stable evolution in ACE2-NEMO, and initial fast increase followed by a sustained slow weakening in ECE3P-control. Total heat flux (panel d) shows larger variability for ACE2-NEMO, but a similar mean value overall, with little trend. Overall, these time series indicate very stable control simulations for ACE2-NEMO.

The pattern of SST and SSH changes are quite different between ACE2-NEMO and ECE3P-control (Fig.~S9), with SSH changes likely driven by modifications made to the air-ice coupling (see Sec.~\ref{sec:sm_shf}). Temperatures in deeper levels of the ocean tend to decrease in ACE2-NEMO, except in the Arctic (Fig.~S10 a), whilst in ECE3P-control the trends are more generalized towards warming (Fig.~S10 b). ACE2-NEMO also shows much stronger surface warming trends than ECE3P-control.

Despite these differences, these results indicate that ACE2 can run stably with unfamiliar boundary conditions provided by NEMO. In turn it can provide NEMO with realistic outputs including fluxes, which drive realistic ocean evolution.

\subsection{Coupled modes of variability}

\begin{figure}[!h]
\centering
    \includegraphics[width=0.9\textwidth]{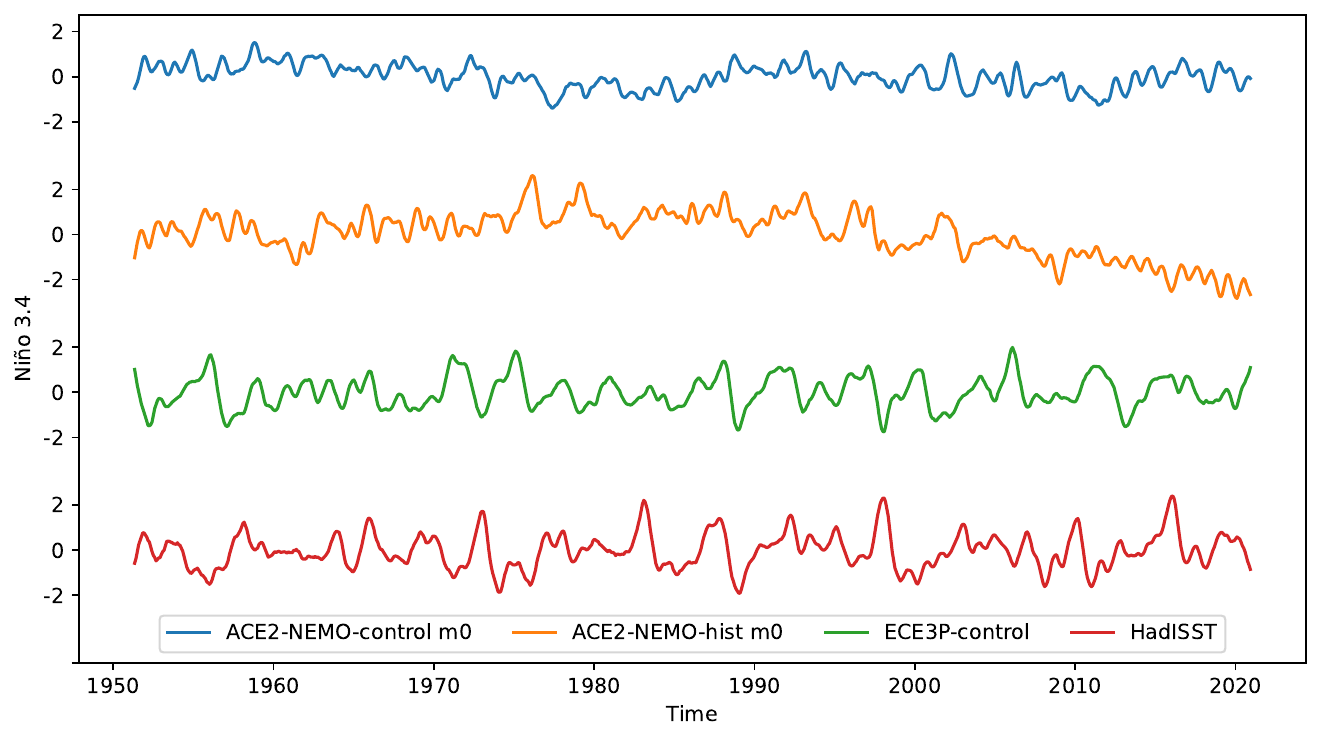}
       \caption{Monthly Ni\~{n}o 3.4 time series for ACE2-NEMO, compared with the ECE3P-control run and HadISST observations. A 5-month rolling average is applied to each time series ACE2-NEMO results show the first ensemble member. }
    \label{fig:nino34_comparison}
\end{figure}

\begin{figure}[!t]
\centering
    \includegraphics[width=\textwidth]{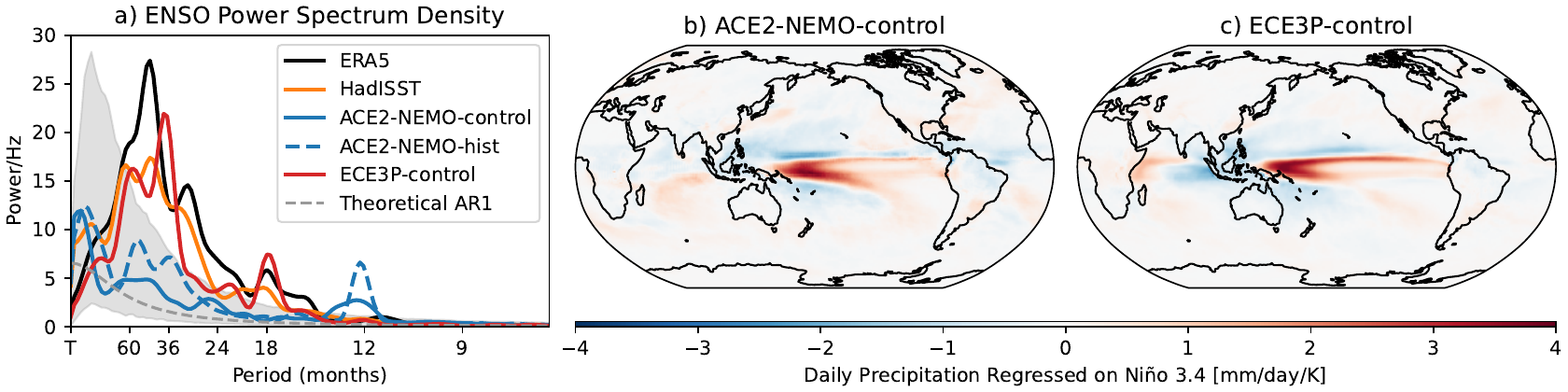}
       \caption{Analysis of ENSO properties of the model simulations (a) Power spectra of the Ni\~{n}o 3.4 index for the different models, ERA5 reanalysis, and HadISST observations. The grey dashed line indicates an AR(1) process fitted to the ACE2-NEMO-control spectra, and the grey shaded area indicates the 2.5\textsuperscript{th}-97.5\textsuperscript{th} percentile range estimated by sampling 1000 AR(1) processes with the same parameters. (b) and (c): Pointwise regression of daily precipitation onto the Ni\~{n}o 3.4 index for the first ensemble member of ACE2-NEMO-control (panel b) and ECE3P-control (panel c).}
    \label{fig:enso}
\end{figure}

\begin{figure}[!t]
\centering
    \includegraphics[width=\textwidth]{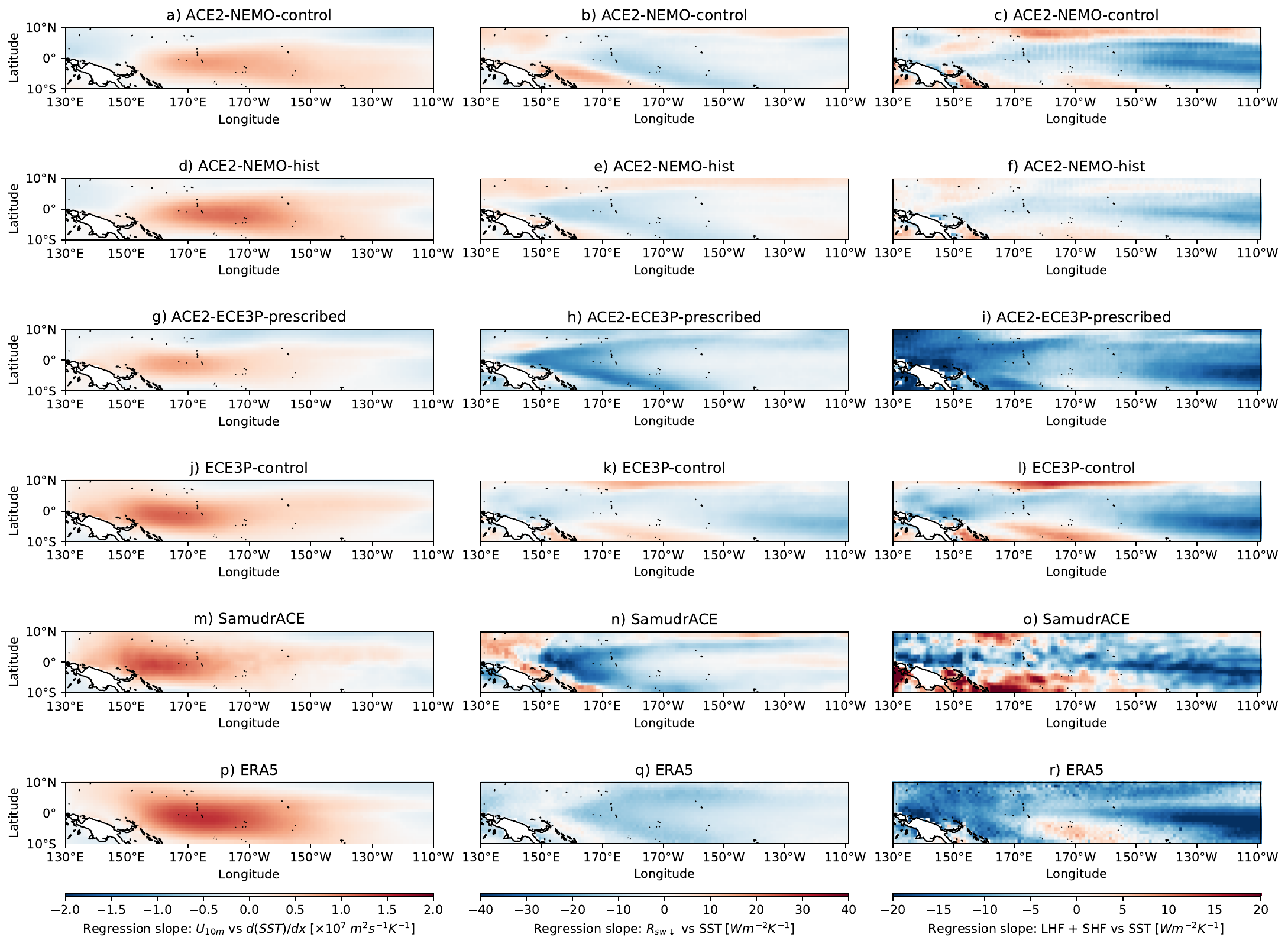}
       \caption{Regression slopes between variables related to ENSO positive and negative feedbacks, organised by column. Left column: 10-metre eastward wind regressed onto domain SST gradient. Middle column: Gridpoint-wise regression of downward short-wave radiation flux on SST. Right column: gridpoint-wise regression of latent plus sensible heat flux on SST. ACE2-NEMO plots use the first ensemble member only.}
    \label{fig:enso_feedback}
\end{figure}

El Ni\~{n}o Southern Oscillation (ENSO) is the key mode of coupled ocean-atmosphere variability \cite{timmermann_nino_2018} and is therefore an important phenomenon to capture accurately in a climate simulation. It also provides a stringent diagnostic of the coupled system, as its dynamics depend sensitively on atmosphere-ocean feedbacks. This makes it a useful test of both the coupling infrastructure and the ability of the atmospheric emulator to generate physically consistent surface fluxes.

The Ni\~{n}o 3.4 series is calculated as the average monthly sea surface temperature anomaly within $5^{\circ}\text{S}-5^{\circ}\text{N}$, $170^{\circ}-120^{\circ}\text{W}$. Monthly anomalies are calculated relative to a climatology calculated as the monthly average over the entire simulation, and anomalies are not de-trended. Plots of the resulting time series are shown in Fig.~\ref{fig:nino34_comparison}.  These time series show that ACE2-NEMO has noticeably different variability in this index, including a reduced magnitude. The effects of the anomalous sea surface behaviour for ACE2-NEMO-hist (see Sec.~\ref{sec:forced_response}) can also be seen after around the year 2000. The oscillations of the index also appear much faster for the ACE2-NEMO runs; possible reasons for this could be the long coupling timestep, which introduces lags that can affect the feedback phase relationships, or that the annual cycle has not been sufficiently removed from the series.

A gridpoint-wise regression of daily precipitation against the Ni\~{n}o 3.4 index is shown in Fig.~\ref{fig:enso} (b) and (c). From this we can see that the response of ACE2 to Pacific SSTs broadly agrees with ECE3P-control, with a realistic pattern of wetting and drying over the tropical Pacific. A notable difference is seen in the Indian Ocean, where ACE2-NEMO shows almost the opposite pattern to ECE3P-control. There are also differences in the eastern tropical Pacific, an increased wetting response in the southern tropical Pacific and a stronger response from ACE2-NEMO in the tropical Atlantic. A similar analysis for ACE2-ECE3P-prescribed (shown in Fig.~S8) shows differences in the precipitation response compared to ACE2-NEMO-control, suggesting a role for the warmer initial conditions in the different response: however, the biases in the ACE2-NEMO-control precipitation response are similar if not smaller than those of the ECE3P-control run when compared to ERA5. Regressions of sea level pressure and surface winds also show broad agreement with ECE3P-control, with biases also particularly notable in the Indian Ocean (Sec.~S9).  

The power spectrum of the Ni\~{n}o 3.4 time series is shown in Fig.~\ref{fig:enso} (a). The computation of spectra is done using the Welch method implemented in scipy (\verb|scipy.stats.welch|) using a Hann window with segment lengths of 256, 128 overlapping points per segment. Time series are de-trended using a degree 2 polynomial, and power spectra are averaged over ensemble members for the ACE2-NEMO runs. 
From this, it is clear that ACE2-NEMO does not capture the temporal variability of ENSO seen in reanalysis, observations, or EC-Earth3P, despite the realistic regression patterns for Ni\~{n}o 3.4. The spectrum for ACE2-NEMO-hist is slightly more peaked around 4 years than that for ACE2-NEMO-control. This may indicate that ACE2 has erroneously learnt an association between ENSO variability and atmospheric forcing. An AR(1) process fitted to the ACE2-NEMO-control data is also shown, including an estimate of the 2.5\textsuperscript{th}-97.5\textsuperscript{th} percentile range. The peaks seen in the ACE2-NEMO-hist run fall within this percentile range, suggesting that the spectra of ACE2-NEMO-control and ACE2-NEMO-hist are not significantly different, and that the ENSO variability in both runs is not distinguishable from a red noise process.

According to the standard theory of how ENSO events are persisted and dispersed, SST anomalies that cause a gradient in temperatures across the central Pacific create surface wind anomalies that positively reinforce the SST anomalies, a process known as Bjerknes feedback \cite{bjerknes_atmospheric_1969, timmermann_nino_2018}. At the same time, surface heat fluxes act to relax the SST anomalies, so that the persistence of ENSO events depends on the balance between these two processes \cite{lloyd_role_2009, xu_revisiting_2025}. To investigate the positive Bjerknes feedback, we calculate the regression slope between monthly surface eastward wind anomalies and the monthly basin-wide Pacific SST gradient, defined as the difference in average temperature between the eastern Pacific ($5^{\circ}\text{S}-5^{\circ}\text{N}$, $140^{\circ}-110^{\circ}\text{W}$) and western Pacific ($5^{\circ}\text{S}-5^{\circ}\text{N}$, $130^{\circ}-160^{\circ}\text{E}$) following \citeA{li_westward_2024}. To investigate the damping effect due to heat fluxes, we calculate pointwise correlations between monthly surface heat flux anomalies and monthly SST anomalies.

The strength of the Bjerknes feedback for ACE2-NEMO-control (Fig.~\ref{fig:enso_feedback} a) is clearly weaker than in ECE3P-control (panel j), SamudrACE (panel m), or ERA5 (panel p). This suggests that the feedback is not strong enough to sustain the SST gradient, leading to a more rapid dissipation of El Ni\~{n}o/La Ni\~{n}a states. A similar explanation can be used to explain the biases seen in the 10-metre winds response to Ni\~{n}o 3.4 (Fig.~S16 c); these may indicate that the shift of the Walker circulation that occurs during El Ni\~{n}o events may not be captured correctly. There also appears to be a slightly stronger feedback for ACE2-NEMO-hist (panel d), which could explain the stronger spectral peaks seen in Fig.~\ref{fig:enso} (a), and adds additional support to the hypothesis that ACE2 has learned to associate some part of the ENSO dynamics with the external forcing. The atmospheric response to SST gradients for ACE2-ECE3P-prescribed are also weak (panel g).

The radiative feedback (Fig.~\ref{fig:enso_feedback}, middle column) shows similarities between ACE2-NEMO-control and ERA5 (panels b and q), but stronger feedback for ACE2-ECE3P-prescribed, ECE3P-control, and SamudrACE (panels h, k, and n). All models except ACE2-ECE3P-prescribed show erroneous positive feedbacks in the northern edge of the domain and around the maritime continent, however the coupled ACE2-NEMO models lack the positive radiative feedback in the eastern tropical Pacific seen for ECE3P-control, a known bias in many climate models \cite{lloyd_role_2012, xu_revisiting_2025}. ACE2-NEMO-control and ACE2-NEMO-hist are broadly similar, with ACE2-NEMO-control showing stronger positive feedback around the maritime continent.

The strength of correlation between heat fluxes and SST (Fig.~\ref{fig:enso_feedback} right column) are more similar between ACE2-NEMO-control and ECE3P-control (panels c and l), with both showing a weaker relationship and erroneous correlations north of the equator compared to ERA5 (panel r). The heat flux response of ACE2-NEMO-hist is also weaker than for ACE2-NEMO-control, which may also contribute to stronger persistence of ENSO states through reduced damping. ACE2-ECE3P-prescribed shows a strong heat flux response to SST, that is perhaps the most similar to ERA5.

Overall, we observe different feedback effects in the coupled ACE2-NEMO models compared to ECE3P-control and ERA5, particularly the strength of the wind response to SST gradients, and the radiative feedback: this remains true for a simulation using prescribed ocean conditions from EC-Earth3P, particularly the wind response to SST gradients, suggesting that this is not merely an artefact of the coupling used here, but the particular reaction of ACE2 to the ocean conditions. The lack of positive wind feedback in particular could explain the difference in Ni\~{n}o 3.4 spectra observed in Fig.~\ref{fig:enso}. There also appear to be differences between the ACE2-NEMO-control and ACE2-NEMO-hist experiments, which could explain the different spectra seen in Fig.~\ref{fig:enso} (a). These findings contrast with the results in \citeA{duncan_samudrace_2025} showing a Ni\~{n}o 3.4 spectrum sharply peaked at a period of around 3 years, and our analysis indicates SamudrACE has realistic feedbacks. The reason for this difference could be due to the additional fine-tuning step used in SamudrACE, which can allow ACE2 to learn appropriate coupled feedbacks, which it perhaps cannot learn from being trained with prescribed sea surface temperatures.

\subsection{Response to forcing}
\label{sec:forced_response}

\begin{figure}[!t]
\centering
    \includegraphics[width=\textwidth]{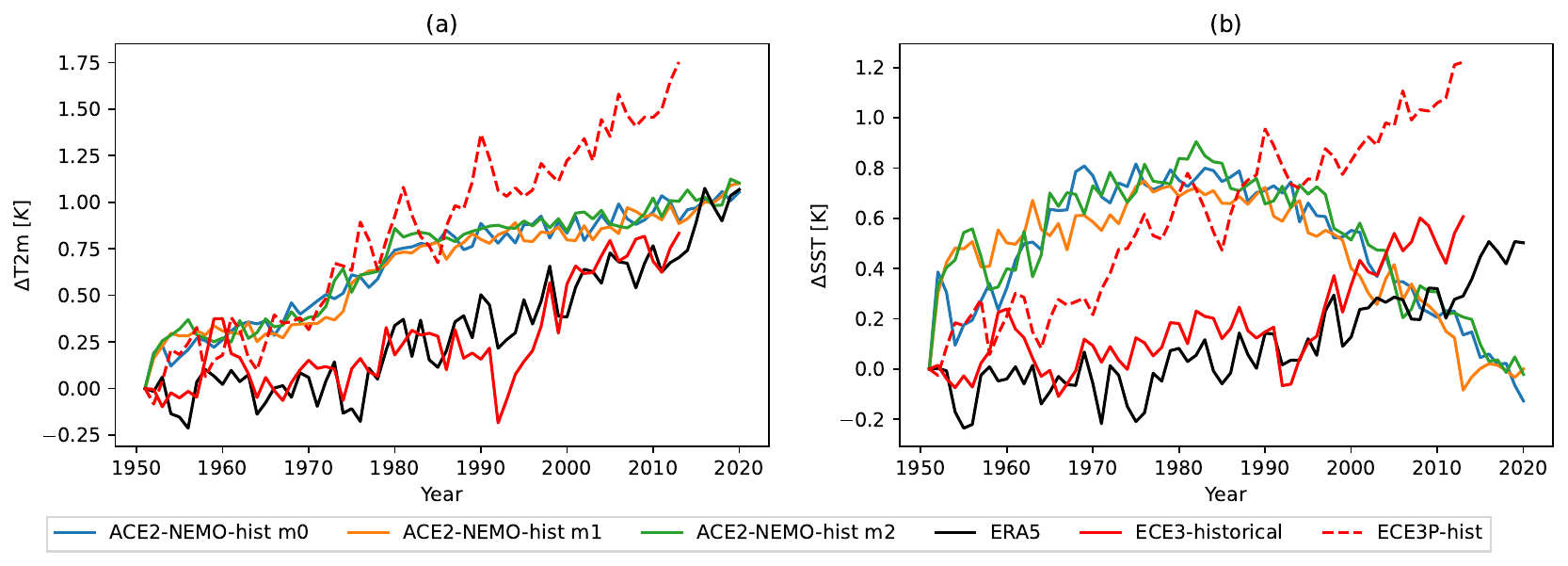}
       \caption{Globally averaged temperatures for the ACE2-NEMO-hist, ECE3P-hist and ECE3-historical experiments, together with ERA5 reanalysis. Temperatures are expressed as changes relative to 1951. Cosine-latitude weighting is used to account for different grid cell size. (a) 2-metre temperature (b) sea surface temperature.}
    \label{fig:historical_temp}
\end{figure}

\begin{figure}[!t]
\centering
    \includegraphics[width=\textwidth]{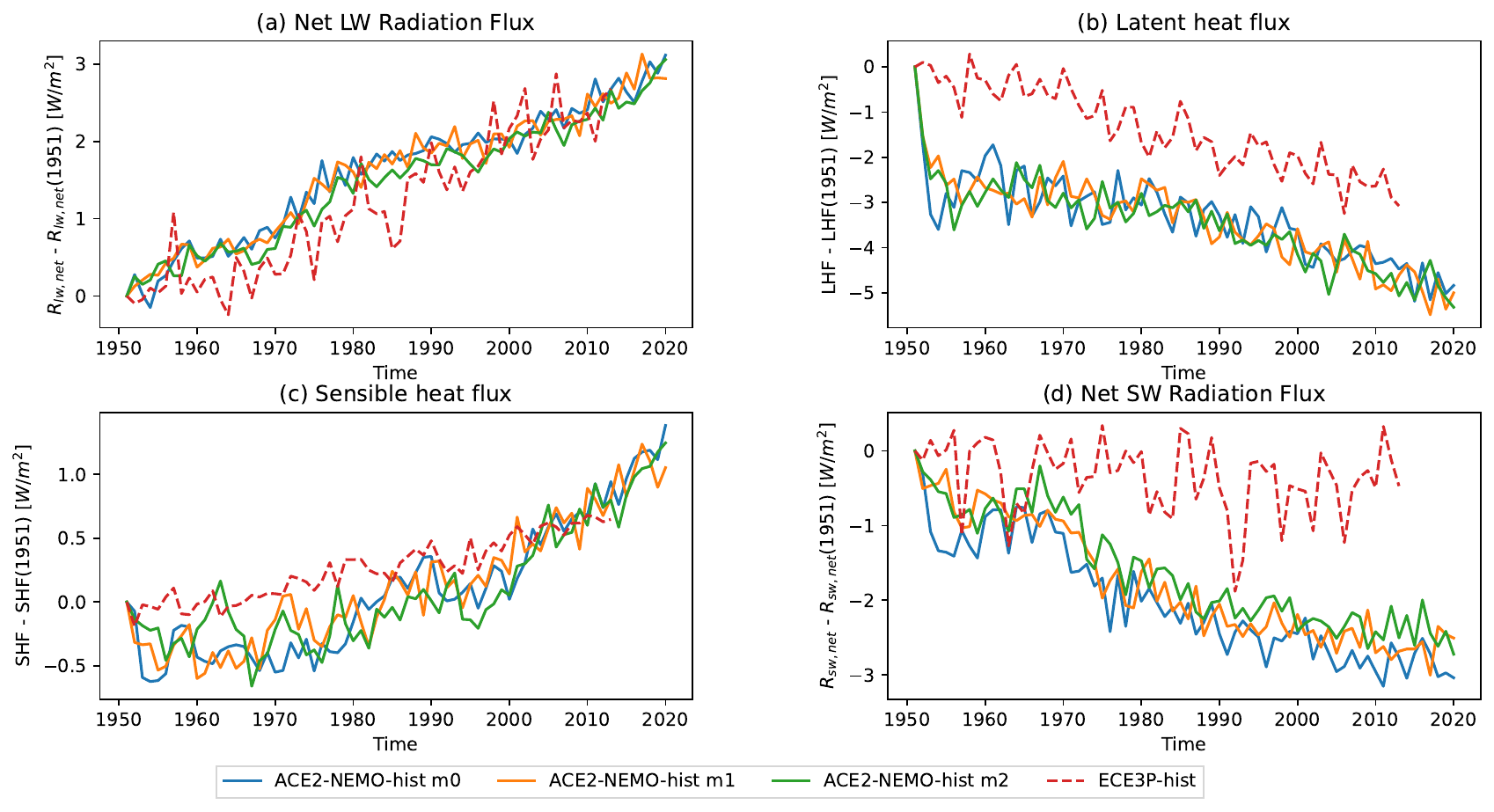}
       \caption{Global averages over ocean points without sea ice for surface heat fluxes. Cosine-latitude weighting is used to account for different grid cell size. (a) net long-wave radiation (b) latent heat (c) sensible and (d) net short-wave radiation heat fluxes. All series are shown relative to the value in 1951.}
    \label{fig:shortwave_historical}
\end{figure}

To evaluate the forced response of ACE2-NEMO, we plot globally averaged 2-metre temperature and SST from the historical simulations in Fig.~\ref{fig:historical_temp}. The evolution of 2-metre temperature in panel (a) shows that the changes of ACE2-NEMO-hist are most consistent with ECE3P-hist up to around 1980, with both showing steeper rises than ECE3-historical and ERA5. This is perhaps to be expected since ECE3P-hist shares the same NEMO configuration as ACE2-NEMO.  

Beyond 1980 however, ACE2-NEMO-hist shows a clear deviation in the 2-metre temperature rise. In contrast to both EC-Earth simulations and ERA5, which show a strong positive temperature trend with an increase of approximately 0.75 degrees over the thirty-year period 1980-2010, ACE2-NEMO shows a surface temperature trend which flattens compared to the earlier period. This coincides with the point at which the ACE2-NEMO-hist SSTs start to deviate from the other series (panel b), with all 3 ensemble members following the same trajectory of decreasing globally averaged temperature (although a 1-day lagged ensemble may not be sufficiently independent to capture the true variability). 

To understand this discrepancy, Fig.~\ref{fig:shortwave_historical} panels a-d show different radiative surface fluxes averaged over all ocean points. While most fluxes show similar changes to ECE3P-historical (Fig.~\ref{fig:shortwave_historical} panels a-c), significant differences in trends are seen in the short-wave radiation fluxes over the ocean, Fig.~\ref{fig:shortwave_historical} (d). This is caused by the downward short-wave fluxes falling at a faster rate than the upward fluxes. Although ACE2 does not explicitly model clouds, this suggests an implicit overestimation of cloud feedbacks over the ocean. It is also notable that ACE2 does not include aerosols as a forcing; given the significant changes in aerosols in the latter half of the 20th century \cite{hoesly2018historical}, this may account for some differences in how ACE2 is behaving in this period. There is also evidence of some redistribution of heat within the ocean, with some increase in temperature seen at depths of 500-1000m (Fig.~S11 b), however this effect seems relatively small and occurs from the year 2000 onwards, much later than the change in SST trend seen in Fig.~\ref{fig:historical_temp} (b).

We note also that there is a rapid increase in the sea surface temperature and 2-metre temperatures observed in the first year in Fig.~\ref{fig:historical_temp}; these occur at the same time as rapid increases in evaporation (see Sec.~S2), which points to a mismatch in temperature gradients between ACE2-NEMO and the ERA5 initial conditions for ACE2, resulting in a short atmospheric `spin-up'.

\section{Discussion}

In this work, we have presented ACE2-NEMO, the first coupling of a pure machine-learnt atmospheric emulator (ACE2) with a dynamical ocean model (NEMO). Apart from changes to the sea ice fluxes, no fine tuning of either component is performed. This creates a unique test of the robustness of the ACE2 atmospheric emulator to out-of-sample inputs, exposes model biases, and provides a novel test of how well the machine learning emulator has learned the appropriate feedbacks to produce realistic coupled variability and response to forcing. It also presents an unexplored approach to integrating machine learning models with dynamical ocean models. Given the relative sparsity of ocean data, integrating with ocean models that model the physical equations may provide an advantage over `full' machine-learning-based ocean emulators in how faithfully they can represent long-term ocean variability.

Achieving the coupling requires overcoming several technical challenges; the models run in different environments, in different programming languages, and the atmospheric model does not produce all of the fluxes required by the ocean. Our solution creates a bridge between the two environments and leverages the existing OASIS coupling infrastructure together with the AirSeaFluxCode library to achieve realistic coupling. Through developing the coupling, several problems were identified relating to the ACE2 fluxes over sea ice; the sensible heat fluxes produced by ACE2 over sea ice did not seem to be consistent with the temperature differences, and there were large drifts in sea ice volume and sea surface height. These problems were solved by using simple bulk formulae to calculate heat fluxes over ice, and enforcing freshwater conservation. The cause of these inconsistencies is not clear; an AMIP-style run using EC-Earth3P sea surface data appears to rule out that this is purely due to the warm ocean initial conditions, and so could be due to out-of-sample inputs from NEMO into ACE2, or because NEMO outputs bulk sea ice temperature rather than skin temperature. In future work we plan to investigate the causes of these inconsistencies more closely, and experiment with different bulk formulae for ice.

The resulting model is stable for at least one hundred years (not shown). We analyse multiple 70-year runs, using fixed 1951 CO\textsubscript{2} forcing (ACE2-NEMO-control) and historical 1951-2021 CO\textsubscript{2} forcing (ACE2-NEMO-hist), compared with similar runs using EC-Earth3 and an AMIP run with prescribed ocean variables. ACE2-NEMO-control shows realistic mean states and fluxes compared to EC-Earth3, confirming that the coupling infrastructure is functioning well. Investigating the coupled El Ni\~{n}o Southern Oscillation (ENSO) phenomenon, we find that, whilst ACE2 variables show realistic correlation patterns with the Ni\~{n}o 3.4 index, the Ni\~{n}o 3.4 spectra for the ACE2-NEMO simulations show a lack of realistic amplitude and temporal variation. Further analysis suggests that this lack of realistic ENSO variability is related to weaker feedback mechanisms in the ACE2-NEMO runs, particularly the response of surface winds to SST gradients, and radiative feedback. This motivates more in-depth analysis of other climate emulators to explore the extent to which the long-timescale ENSO variability is driven by correctly learned atmospheric feedbacks. The difference in spectrum for the control and historical runs also suggests that ACE2 may have learnt to associate CO\textsubscript{2} forcing with ENSO variability. Since we use the version of ACE2 trained on ERA5, this may be due to over-fitting to a short number of ENSO events under global warming conditions, and it may be that using a version of ACE2 trained on a longer climate model run would produce more realistic variability. It also raises the question of whether future ML emulators should be trained to produce forced responses and variability separately. It is also possible that the lack of stochasticity in ACE2 contributes to the incorrect variability \cite{christensen_stochastic_2017}.

The ACE2 6-hour model step is substantially longer than the coupling time step of around 45 minutes for EC-Earth3 with similar atmospheric and ocean resolutions \cite{acosta_balancing_2023}. This relatively long coupling step means that NEMO sees fluxes that can be up to 6-hours delayed, with potential effects on feedback relationships and coupled variability. This is mitigated to some extent by the fact that fluxes calculated by ACE2 are output as 6-hour averages (although other fluxes such as the ice fluxes are calculated using instantaneous variables). One method to reduce the coupling timestep would be to use time interpolation; since ACE2 takes very little time to calculate a 6-hour timestep compared to NEMO, then the prediction made by ACE2 six hours ahead can be combined with the current ocean surface conditions to estimate future fluxes, allowing more granular fluxes to be calculated as a linear interpolation between the present and future estimated fluxes. Within our current setup however, this requires an extra flux calculation step, which would slow the simulation down. Conversely, it is also possible that some of the drifts and biases present in ACE2-NEMO are in fact due to unfiltered high-frequency errors in ACE2 fluxes being passed to NEMO, and that averaging fluxes over longer timesteps (as done for SamudrACE) could act to remove these biases. We leave this exploration of the effect of the coupling timestep for future work.

The ACE2-NEMO-hist run also shows that, whilst ACE2-NEMO shows a global increase in 2-metre temperature that is initially in line with EC-Earth 3, significant sea surface cooling occurs from 1980 onwards, which appears to be due to an incorrect response of downward short-wave radiation to the forcing, although it is also possible that the lack of aerosol input to ACE2 plays a role in how well the model simulates cloud behaviour. Apart from highlighting a potential sensitivity within ACE2, this also highlights the need for additional fields (such as cloud fraction) to be produced by climate emulators in order to facilitate diagnosis and improve interpretability. 

We note some limitations of this study. The initial conditions used in this paper are warmer than the ERA5 initial conditions for the same period, by around 1K. Previous results on ACE2 running with perturbed oceans indicate that this may be an additional source of error in ACE2, and so some of the effects noted above may be due to this \cite{zhang_equilibrium_2025}. Whilst our AMIP simulation has shed some light on the effects of this, a useful future experiment would start from ocean initial conditions that are closer to the reanalysis, perhaps via a nudged spin-up of the ocean. The efficiency of the model is limited by the flux calculations, since it is not optimised for this scale of data; significant speed-up of this module would therefore be expected with the appropriate optimisation. Additionally, more recent iterations of ACE2 such as that used in \citeA{duncan_samudrace_2025} now include momentum fluxes, removing most of the need for a separate flux calculation. To ensure stability of the model, heat fluxes over ice near coastal points were masked: this is undesirable, as it affects the polar heat budgets, and may have other effects on the model biases. It is worth noting that the developers of SamudrACE encountered similar problems, attributed to the effects of regridding, and inconsistent masks, which were removed by fine tuning \cite{duncan_samudrace_2025}. The coupled model may therefore be improved by addressing inconsistencies between the land/sea masks of NEMO and ACE2 and investigating the effects that regridding has on the fluxes. NEMO and ACE2 also have different definitions of sea ice fraction; for NEMO it is defined as the fraction of ocean, whilst for ACE2 it is defined as the fraction of a grid cell. Whilst the code that passes fluxes from ACE2 to NEMO uses the NEMO sea ice fraction, passing ACE2 the unmodified NEMO sea ice fraction as input could be one source of the instability. Freshwater conservation has been enforced, to address the imbalance between the freshwater fluxes from ACE2 and the climatological surface runoff. Investigating ways to address this imbalance, such as including a surface runoff model, as well as investigating global heat conservation, is another key area of model improvement that increases physical realism and may remove biases. To provide solid precipitation as an input to NEMO, a simple heuristic model was used; there is plenty of scope to use a more sophisticated approximation that reduces solid precipitation biases.

There are many interesting future extensions of this work. The same coupling framework could be extended to other atmospheric emulators, such as NeuralGCM \cite{kochkov_neural_2024}, as a unique test of realism and robustness that extends previous tests using prescribed SSTs \cite{zhang_equilibrium_2025}. Given recent results demonstrating skilful seasonal forecasts using machine learning \cite{kent_skilful_2025, antonio_seasonal_2026, zhang_advancing_2025} it would be insightful to run coupled seasonal forecasting using ACE2 and other models. Since the computational resources are reduced compared to a full coupled model, there is also potential to apply this framework to achieve more efficient coupled ocean spin-ups, which can improve the accuracy of climate simulations \cite{gupta2013climate}; this will be explored in forthcoming work.

\section{Data availability}

 Processed data used to produce the plots in this manuscript are available in an accompanying Zenodo repository \cite{Antonio_ACE2NEMO_2026}. ACE2 initial conditions are taken from the pre-prepared initial conditions detailed in \citeA{watt-meyer_ace2_2025}, SamudrACE initial conditions are taken from the Hugging Face model repository \cite{allenai_samudrace_cm4_picontrol}. The ERA5 datasets used are available in the Copernicus Climate Change Service, Climate Data Store, (2023), for registered users: the dataset used is the ERA5 hourly data on single levels from 1940 to present  \cite{Hersbach_ERA5hourly_2023}. The historical EC-Earth3 simulation is presented in \citeA{bilbao_assessment_2020} and \citeA{ecearth_historical_2019}. The additional EC-Earth3 simulations from PRIMAVERA are presented in \citeA{haarsma_highresmip_2020} and \citeA{moreno2025very}. HadISST data is detailed in \citeA{rayner2003global}.

\section{Code availability}
% TODO: insert zenodo record when approved
The ACE2-NEMO coupling code and the code used to analyse the data is available at  \url{https://github.com/bobbyantonio/ace2_nemo_coupler}. This relies on a forked version of the EC-Earth4 repository, which is available at \url{https://git.smhi.se/e8118/ecearth4} after registration (see \url{https://ec-earth-4-docs.readthedocs.io/en/latest/}).The modified version of ACE2 can be found in the forked repository at \url{https://github.com/bobbyantonio/ace}. The SamudrACE simulation was performed using the code at \url{https://github.com/ai2cm/ace}.

\section*{Acknowledgements}

 The authors are grateful to the Editors, two anonymous reviewers and Laure Zanna for their insightful comments which greatly improved the manuscript. This publication is part of the EERIE project funded by the European Union (Grant Agreement No 101081383). Views and opinions expressed are however those of the author(s) only and do not necessarily reflect those of the European Union or the European Climate Infrastructure and Environment Executive Agency (CINEA). Neither the European Union nor the granting authority can be held responsible for them. This work was funded by UK Research and Innovation (UKRI) under the UK government’s Horizon Europe funding guarantee (grant number 10049639). HMC was also funded through a Leverhulme Trust Research Leadership Award `Seamless Uncertainty Quantification for Earth System prediction' (SUQCES). For the purpose of Open Access, the author has applied a CC BY public copyright licence to any Author Accepted Manuscript version arising from this submission.

\section*{Author Contributions}
BA, KS and HC contributed to the study conception and design. Material preparation, data generation and analysis were performed by BA. PO provided the EC-Earth 3 data and ocean initial conditions and advised on ocean model configuration. The first draft of the manuscript was written by BA and all authors commented on subsequent versions of the manuscript. All authors read and approved the final manuscript.

\section*{Conflicts of Interest}

The authors declare there are no conflicts of interest for this manuscript.

\bibliography{acenemo_references}

\newpage

\section*{Supplementary materials}

%%%%
\renewcommand{\thesection}{S\arabic{section}}
\renewcommand{\thefigure}{S\arabic{figure}}
\setcounter{section}{0}
\setcounter{figure}{0}

\raggedbottom
%%\unnumbered% uncomment this for unnumbered level heads

\section{Coupling details}

\begin{figure}
% \setfigurenum{S1} %%You can change number for each figure if you want, not required. "S" prepended automatically.
\noindent\includegraphics[width=\textwidth]{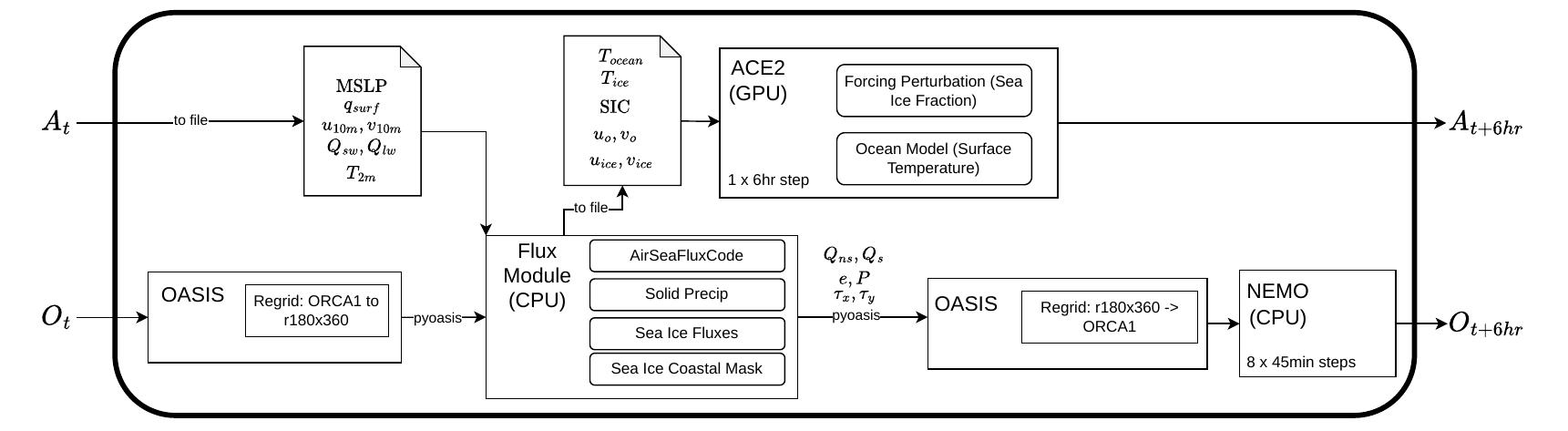}
\caption{Detailed diagram of the coupling between ACE2 and NEMO. Symbols are defined in the text of Sec. S1.}
\label{fig:detailed_coupling}
\end{figure}

Fig.~\ref{fig:detailed_coupling} shows a detailed diagram of how ACE2 and NEMO exchange variables and fluxes, mediated by the flux calculation module. Symbols in this diagram are defined as follows: mean sea level pressure (MSLP), surface humidity ($q_{surf}$), 10-metre winds ($u_{10m}, v_{10m}$), net surface short-wave radiation ($Q_{sw}$), net surface long-wave radiation ($Q_{lw}$), 2-metre temperature ($T_{2m}$), sea surface temperature ($T_{ocean}$), sea ice bulk temperature ($T_{ice}$), sea ice fraction ($f_{ice}$), ocean surface velocities ($u_o, v_o$), sea ice velocities ($u_{ice}, v_{ice}$), non-solar heat flux ($Q_{ns}$), solar heat flux ($Q_s$), evaporation ($e$), precipitation ($P$), momentum fluxes ($\tau_x, \tau_y$).

Table \ref{tab:variable_summary} provides a summary of how different fluxes are calculated and regridded, as well as the time period over which they are defined (either a 6-hour mean, or an instantaneous snapshot). The source column indicates whether the flux calculation uses raw ACE2 values, is calculated using AirSeaFluxCode (with ACE2 and NEMO surface state variables as input), using the CORE bulk formulae (with ACE2 and NEMO surface state variables as input), or via the solid precipitation heuristic. 

\begin{table}[h!]
\centering
\begin{tabular}{ c || c | c | c | c }
 Flux quantity &  Units & Time & Source & Regrid type\\ 
 \hline 
 \hline
 Latent heat flux (ocean) & ${W m}^{-2}$ & 6hr mean & ACE2 & Bilinear\\  
  Latent heat flux (ice) & ${W m}^{-2}$ & Snapshot & CORE Bulk Formulae &Bilinear\\  
 Sensible heat flux (ocean) & ${W m}^{-2}$  & 6hr mean  & ACE2  &Bilinear\\
 Sensible heat flux (ice) & ${W m}^{-2}$  & Snapshot  &  CORE Bulk Formulae  &Bilinear\\
 Evaporation (ocean)  & $kg m^{-1}s^{-1}$  & 6hr mean  & AirSeaFluxCode &Bilinear\\
 Evaporation (ice)  & $kg m^{-1}s^{-1}$  & Snapshot  & CORE Bulk Formulae & Conservative\\
 Liquid Precipitation & $kg m^{-1}s^{-1}$ & 6hr mean & ACE2 & Conservative\\
 Solid Precipitation & $kg m^{-1}s^{-1}$ & 6hr mean & Solid precip heuristic &Conservative\\
 Momentum (ocean) &$kg m^{-1}s^{-2}$ & Snapshot & AirSeaFluxCode &Bilinear\\
 Momentum (ice) &$kg m^{-1}s^{-2}$ & Snapshot & CORE bulk formulae &Bilinear\\
 SW Radiation & ${W m}^{-2}$& 6hr mean& ACE2 &Bilinear\\
 LW Radiation & ${W m}^{-2}$& 6hr mean& ACE2 &Bilinear \\
\end{tabular}
 
\caption{Details of the flux variables passed to NEMO.}
\label{tab:variable_summary}
\end{table}

\section{Sea surface temperature behaviour}
\label{sec:sstbehaviour}

\begin{figure}[!t]
\centering
    \includegraphics[width=0.9\textwidth]{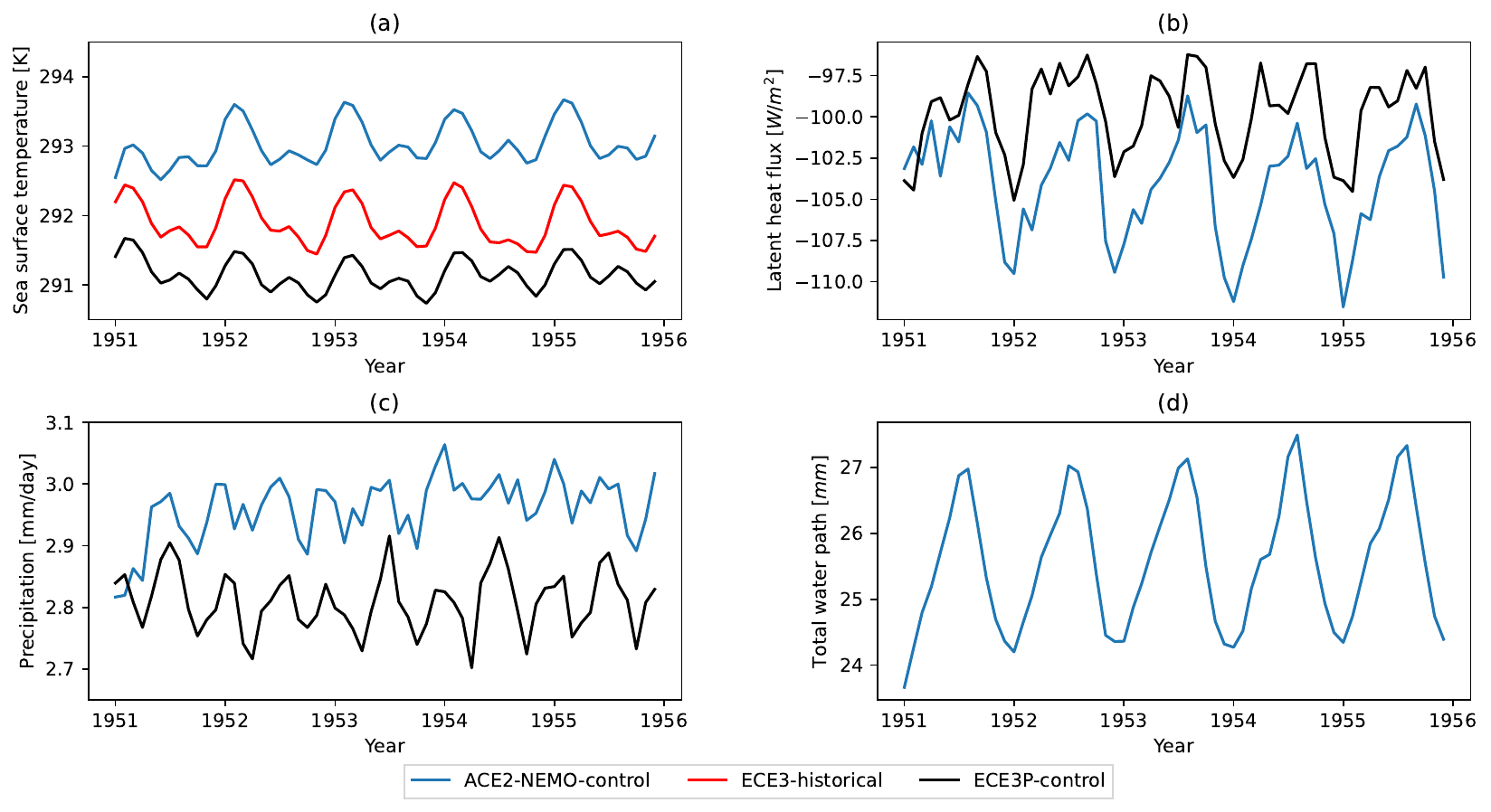}
       \caption{A comparison of the evolution of (a) SST  (b) latent heat flux (c) daily precipitation and (c) total water path for the first five years of the ACE2-NEMO-control experiment (first ensemble member only), compared to ECE3-historical and ECE3P-control.}
    \label{fig:sst_comparison}
\end{figure}

In this section, we investigate the changes in sea surface temperature during the control and historical runs.
Firstly, we investigate the atmosphere `spin-up' effects, which results in the sea surface temperature rising rapidly within the first simulation year. Plots of several globally averaged variables are shown in Fig.~\ref{fig:sst_comparison}. Panel a in this figure shows the change in global SST; firstly this demonstrates the $\sim 0.5$K SST difference between ECE3P-control and ECE3-historical. ACE2-NEMO-control starts from the same initial conditions as ECE3-historical, but quickly moves to a new equilibrium around 1K warmer than ECE3-historical. Panel b demonstrates that the latent heat flux decreases over the first year, reaching a new equilibrium of lower latent heat fluxes (i.e. more cooling), indicating increased evaporation. Panel c demonstrates that during this initial period, there is a large increase in precipitation, and panel d demonstrates that subsequent Januarys have increased total water path relative to the first month, consistent with the increase in evaporation. It therefore appears that for the first year, the air-sea temperature gradients are larger than ACE2's initial conditions, such that the level of evaporation is too low, and other sources of ocean heating are not balanced by cooling from evaporation.

\section{Mean fluxes}
\label{sec:mean_state_fluxes}

\begin{figure}[!t]
\centering
    \includegraphics[width=0.8\textwidth]{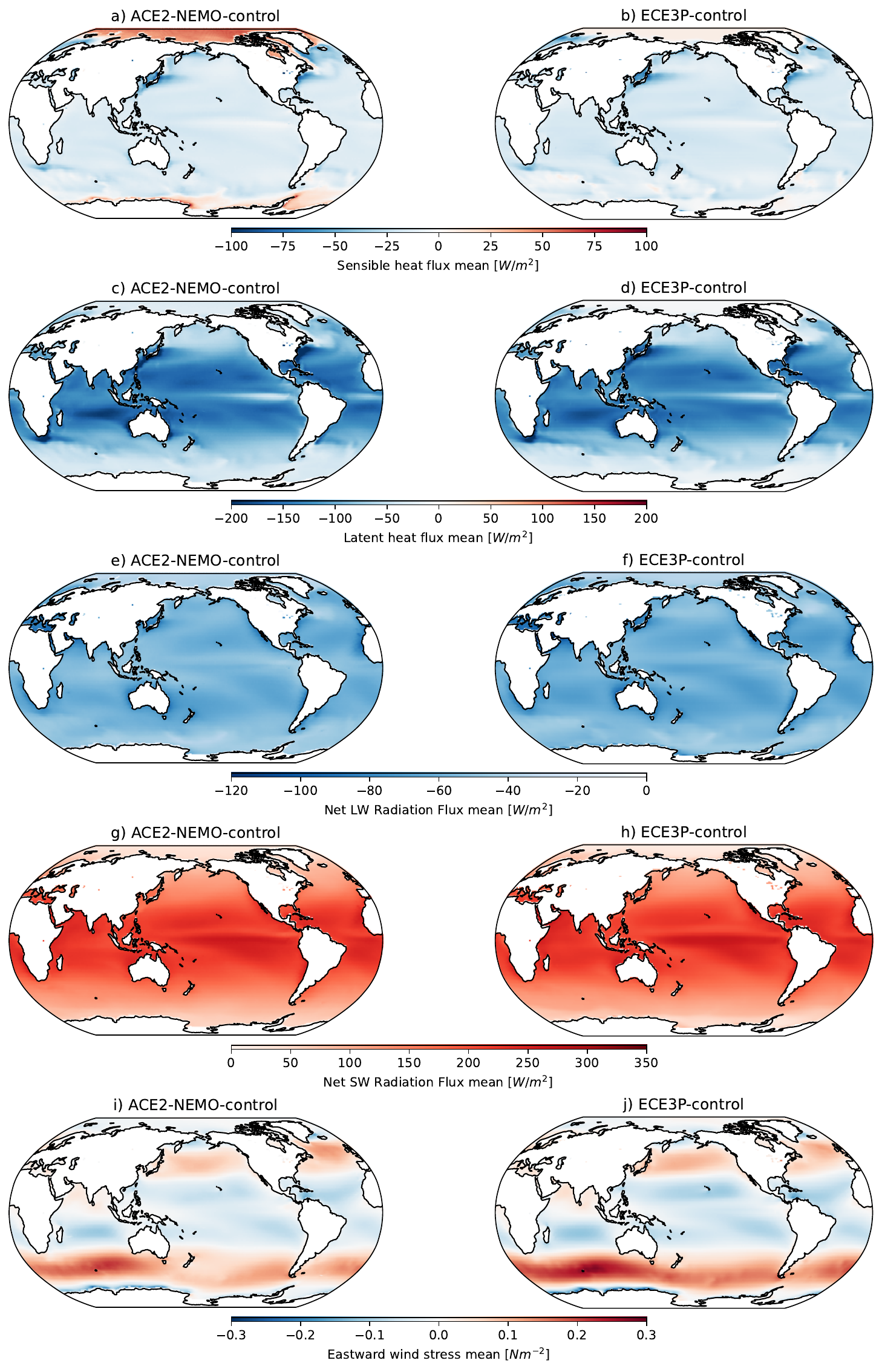}
       \caption{Mean fluxes for ACE2-NEMO-control (first ensemble member) compared to ECE3P-control: (a) and (b): sensible heat flux, (c) and (d): latent heat flux, (e) and (f): net surface long-wave radiation flux, (g) and (h): net surface short-wave radiation flux, (i) and (j) Eastward wind stress. Positive fluxes are downwards.}
    \label{fig:mean_state_fluxes}
\end{figure}

Heat and freshwater fluxes averaged over the entire 70-year ACE2-NEMO-control run are shown in Fig.~\ref{fig:mean_state_fluxes}. The fluxes follow the convention in NEMO, that a positive value indicates a gain of that quantity for the ocean.

\section{Seasonal cycle and interannual variability}

\begin{figure}[!t]
\centering
    \includegraphics[width=0.9\textwidth]{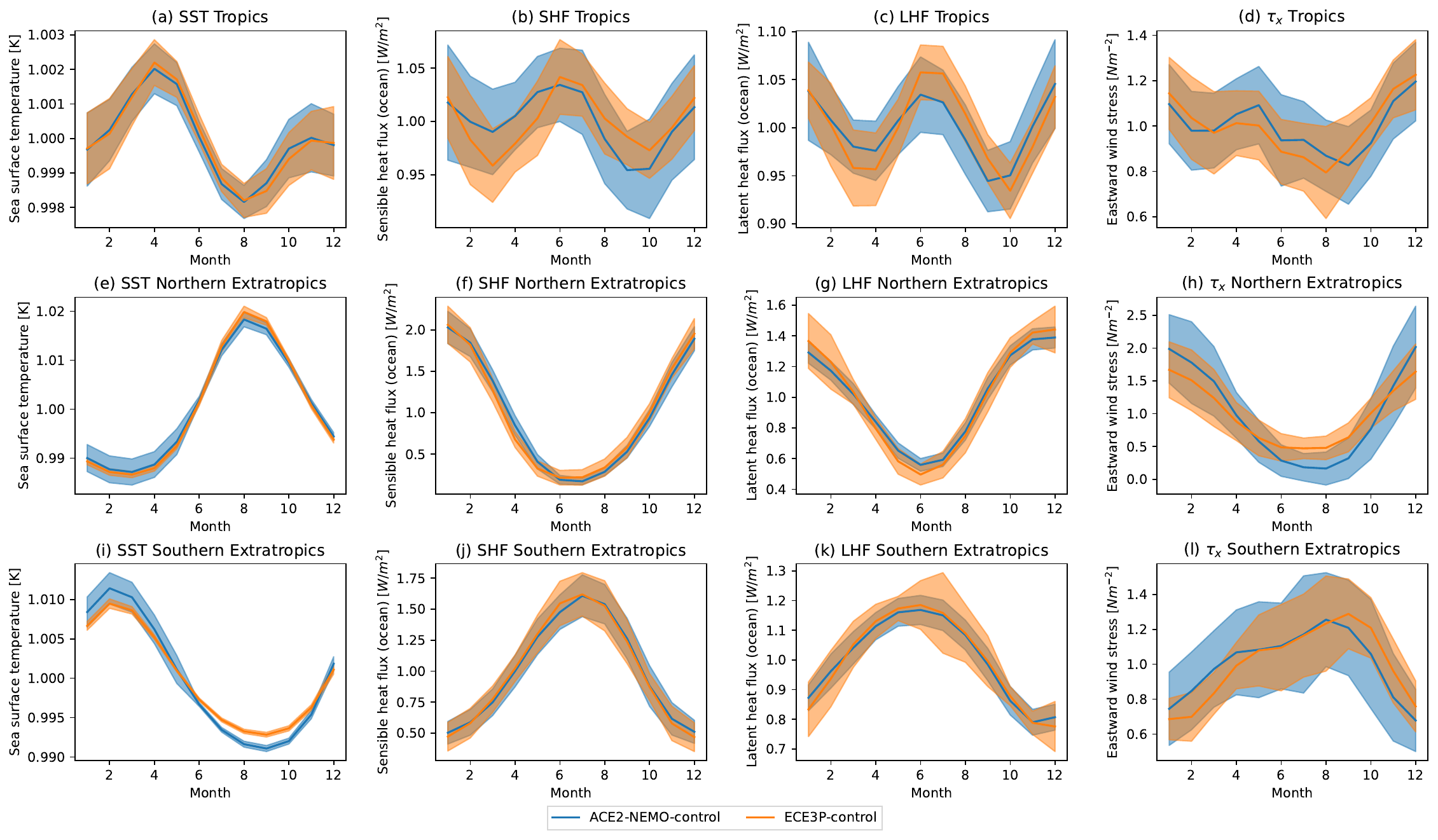}
       \caption{Seasonal cycle of variables from ACE2-NEMO-control (first ensemble member) and ECE3P-control, relative to the mean of each year ( see main text for details of how these values are calculated). Shading indicates 2 standard deviations either side of the monthly mean. Top row: tropics, Middle row: northern extratropics, Bottom row: southern extratropics.}
    \label{fig:seasonal_cycle}
\end{figure}

The seasonal cycle of sea surface temperature, sensible heat flux, latent heat flux, and eastward momentum flux are shown in Fig.~\ref{fig:seasonal_cycle}. These are calculated by first calculating the average monthly value for each year, and dividing the monthly values by the average for their respective year. This then gives a representation of the seasonal cycle excluding interannual variability. Shading indicates 2 standard deviations either side of the mean seasonal cycle. Results are shown over sea points only, and are split into three regions: Tropics (20\textsuperscript{o}S-20\textsuperscript{o}N), northern extratropics (30\textsuperscript{o}-70\textsuperscript{o}N), and southern extratropics (70\textsuperscript{o}-30\textsuperscript{o}S). Overall we can see good agreement between the annual cycles of ACE2-NEMO-control and ECE3P-control, with similar levels of variance. The seasonal cycle for sea surface temperature in the extratropics (panels e and i) shows lower variance for ECE3P-control, and covers a smaller range in the southern extratropics, possibly related to the large variability of the sea ice (Sec.~\ref{sec:sea_ice}). Sensible and latent heat fluxes in the tropics also show differences in the relative strengths of fluxes in March and October.

\begin{figure}[h]
\centering
    \includegraphics[width=0.9\textwidth]{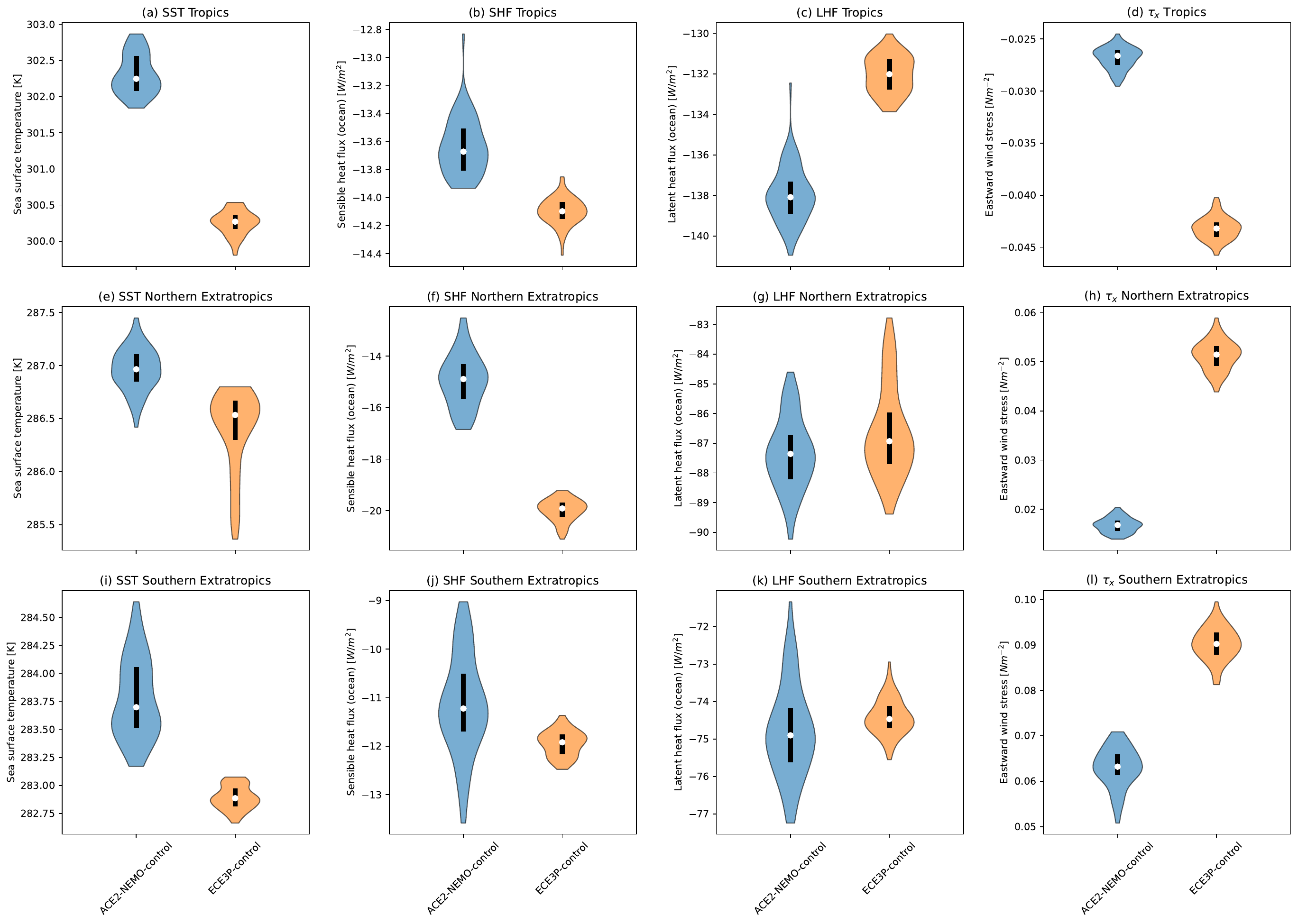}
       \caption{Interannual variability of variables from ACE2-NEMO-control (first ensemble member) and ECE3P-control. Median values shown as white dots, and interquartile ranges shown as the thick black lines. See main text for a description of how these values are calculated. Top row: tropics, Middle row: northern extratropics, Bottom row: southern extratropics.}
    \label{fig:interannual_variability}
\end{figure}

Plots of the interannual variability for the same variables are shown in Fig.~\ref{fig:interannual_variability}. These show the distribution of the annual mean of monthly values, with median values shown as white dots, and interquartile ranges shown as the thick black lines. Most of the data points for ACE2-NEMO-control and ECE3P-control have similar magnitudes of ranges, albeit with noticeable offsets in the median values. Sea surface temperature (panels a, e, and i) is consistently higher for ACE2-NEMO-control than ECE3P-control, as expected from the warmer ocean initial conditions used by ACE2-NEMO. Interannual variability in the southern extratropics is particularly high, perhaps because of the larger variability in sea ice due to the use of bulk formulae in the air-ice fluxes (Sec.~\ref{sec:sm_shf}). Sensible and latent heat fluxes in the extratropics (panels f, g, j, and k) are consistently lower for ACE2-NEMO-control in the extratropics, and generally show a greater range of values over all areas. The momentum fluxes (panels d, h, and l) show similar ranges of interannual variability, but much higher fluxes in the tropics (panel d) whilst much lower fluxes in the extratropics (h and l). 

\section{Sensitivity experiments}
\label{sec:sensitivity}
\begin{figure}[!t]
\centering
    \includegraphics[width=0.9\textwidth]{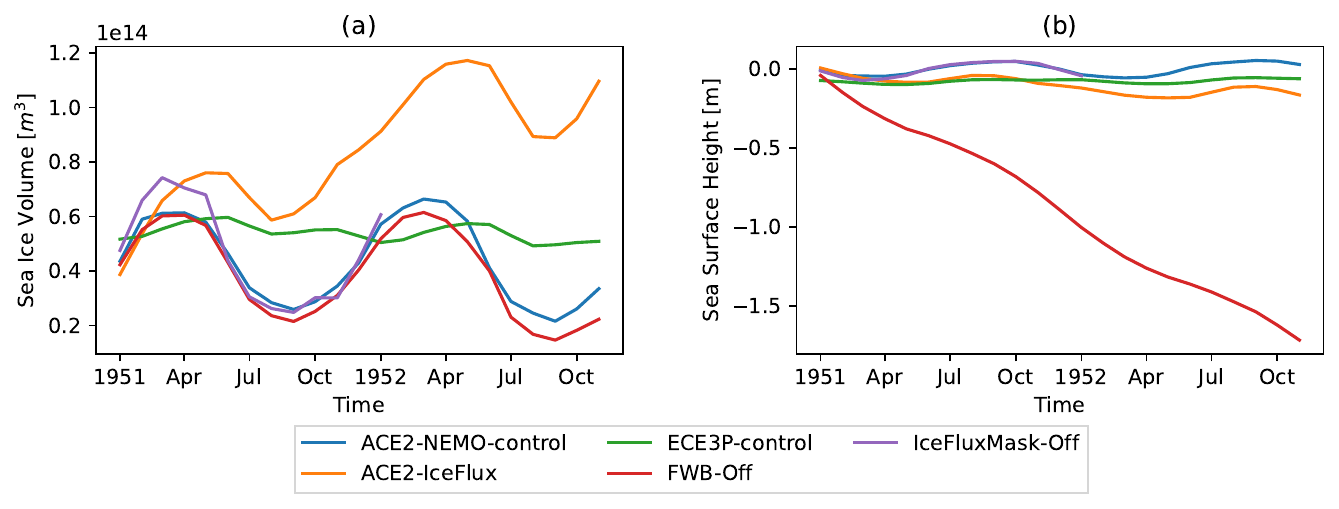}
       \caption{Evolution of (a) sea ice volume and (b) sea surface height for an ACE2-NEMO simulation using the fluxes from ACE2 over sea ice points instead of fluxes calculated using the CORE bulk formulae.}
    \label{fig:ace2_drift}
\end{figure}

In order to assess the impacts of the different modelling fixes we have made in order to stably couple ACE2 and NEMO (namely the sea surface freshwater conservation enforcement, CORE ice bulk formulae, and masking of coastal ice fluxes), we perform several sensitivity experiments with these fixes removed. 

The evolution in global sea surface height and sea ice volume is shown in Fig.~\ref{fig:ace2_drift}, comparing the ACE-NEMO-control run with several short runs that test how different modelling choices affect these variables. ACE2-IceFlux denotes a simulation using the native fluxes produced by ACE2 instead of the CORE bulk formulae. IceFluxMask-Off indicates a simulation without the coastal ice flux masking (for this reason this simulation was only stable for just over 1 year). FWB-Off indicates a simulation where the enforced freshwater balance in NEMO is switched off.

In Fig.~\ref{fig:ace2_drift} (a), we can see the effect that these adjustments have on the sea ice volume. Using the ACE2 ice fluxes, we see a steady increase in global sea ice volume. The simulation without coastal ice flux masking shows some increase in ice volume around February-June, but similar volumes in the other months. 

In Fig.~\ref{fig:ace2_drift} (b), we can see the effect that these adjustments have on the sea surface height. The most significant effect is the enforcement of the freshwater conservation at the ocean surface is switched off; this results in a sea surface height drift of around -1m per year. Fig.~\ref{fig:ace2_drift} (b) also shows that using ACE2 ice fluxes results in a sea surface height drift, although much smaller at around -0.1m per year.

\section{Effect of warm ocean initial conditions}

\begin{figure}[!t]
\centering
    \includegraphics[width=0.9\textwidth]{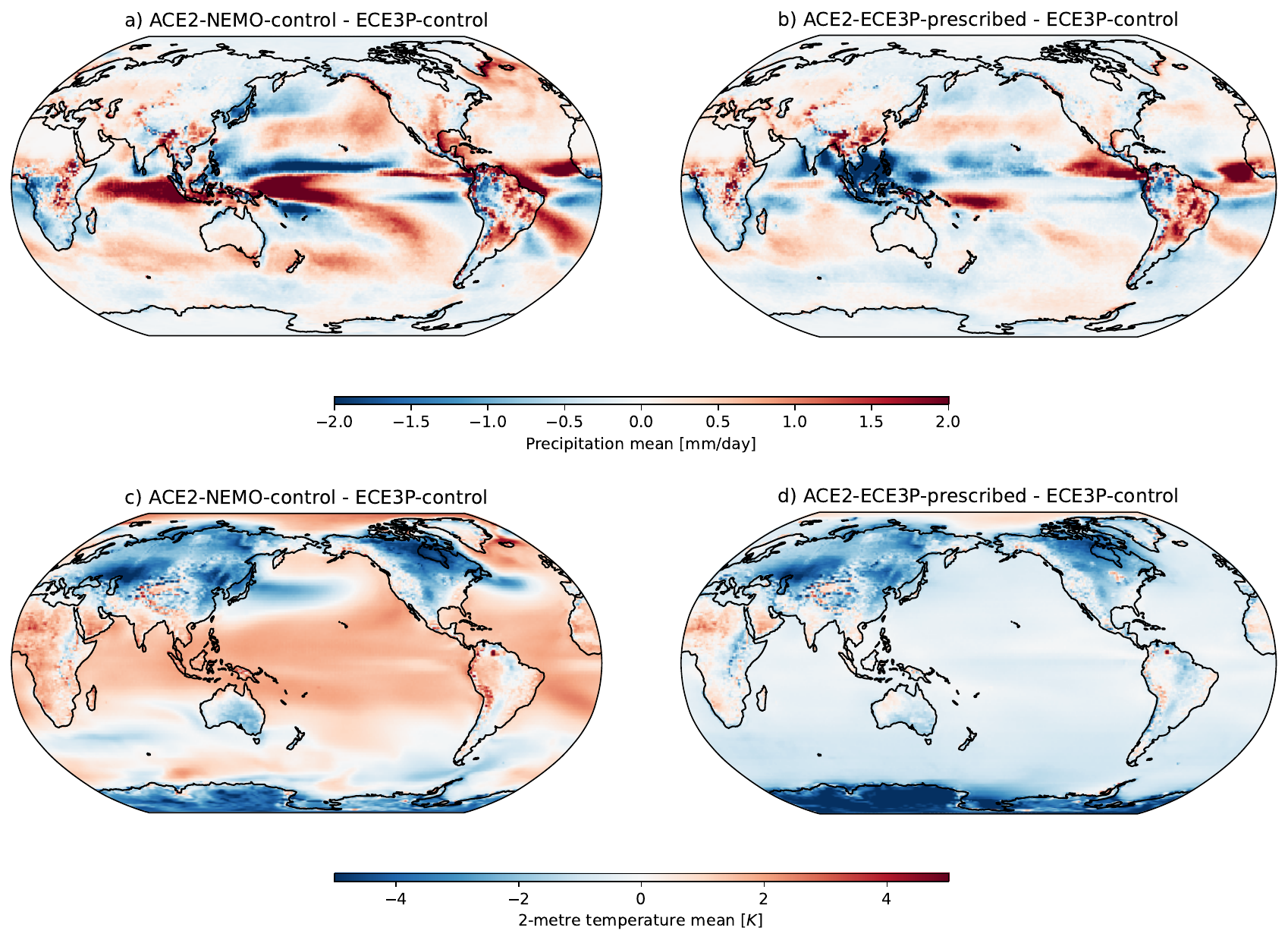}
       \caption{Mean state biases for precipitation (top row) and 2-metre temperature (bottom row) for ACE2-NEMO-control (left column) and ACE2-ECE3P-prescribed (right column). Bias is calculated relative to ECE3P-control.}
    \label{fig:mean_state_biases_comparison}
\end{figure}

\begin{figure}[!t]
\centering
    \includegraphics[width=\textwidth]{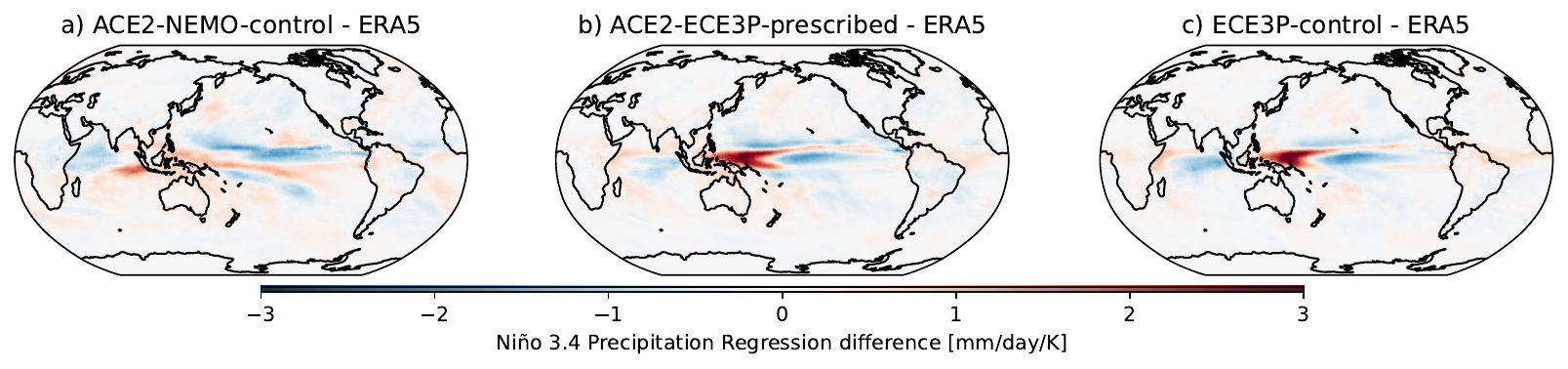}
       \caption{Plots showing the bias in precipitation response to Ni\~{n}o 3.4 relative to ERA5 a) ACE2-NEMO-control, b) ACE2-ECE3P-prescribed, and c) ECE3P-control.}
    \label{fig:enso_regression_bias}
\end{figure}

% \begin{figure}[!t]
% \centering
%     \includegraphics[width=0.9\textwidth]{enso_regression_comparison_era5.pdf}
%        \caption{As in fig.~\ref{fig:enso_regression_bias}, but comparing to ERA5 instead of ECE3P-control.}
%     \label{fig:enso_regression_bias_era5}
% \end{figure}

In order to estimate the effects of the warm NEMO initial conditions, we include additional analysis of the simulation using prescribed ECE3P ocean variables, ACE2-ECE3P-prescribed (see Sec.~\ref{sec:data}). Fig.~\ref{fig:mean_state_biases_comparison} compares the mean state biases of ACE2-NEMO-control and ACE2-ECE3P-prescribed to ECE3P-control; for precipitation (panels a and b), we see significantly more wetting in the tropics, particularly in the western Pacific and Indian Ocean. There are also areas of reduced precipitation in northern and southern tropical Pacific, and some smaller differences observable in the mid-latitudes, producing predominantly wetter conditions. For 2-metre temperature (panels c and d), ACE2-NEMO-control shows a consistent warm bias over ocean points, whilst ACE2-ECE3P-prescribed shows biases of overall smaller magnitude, in the opposite direction. This is an intuitive outcome given the much warmer sea surface temperatures used as initial conditions in the ACE2-NEMO-control run.

Fig.~\ref{fig:enso_regression_bias} compares the differences in precipitation responses to Ni\~{n}o 3.4, compared to ERA5 reanalysis. The prescribed SST AMIP simulation (panel b) generally produces biases in precipitation response that is more similar to ECE3P-control (panel c), as is to be expected, with a particularly significant bias over the maritime continent. The ACE2-NEMO-control run shows a reduced response in the northern tropical Pacific, and there is also increased response extending South East from the maritime continent, along the South Pacific convergence zone. There is also a significant area of anomalous response seen in the Indian Ocean for ACE2-NEMO-control, which also appears to persist in the prescribed SST simulation. The overall bias in precipitation response is of similar magnitude or smaller than the biases for ECE3P-control and ACE2-ECE3P-prescribed.

\section{Model Drift}

\begin{figure}[!h]
\centering
    \includegraphics[width=0.9\textwidth]{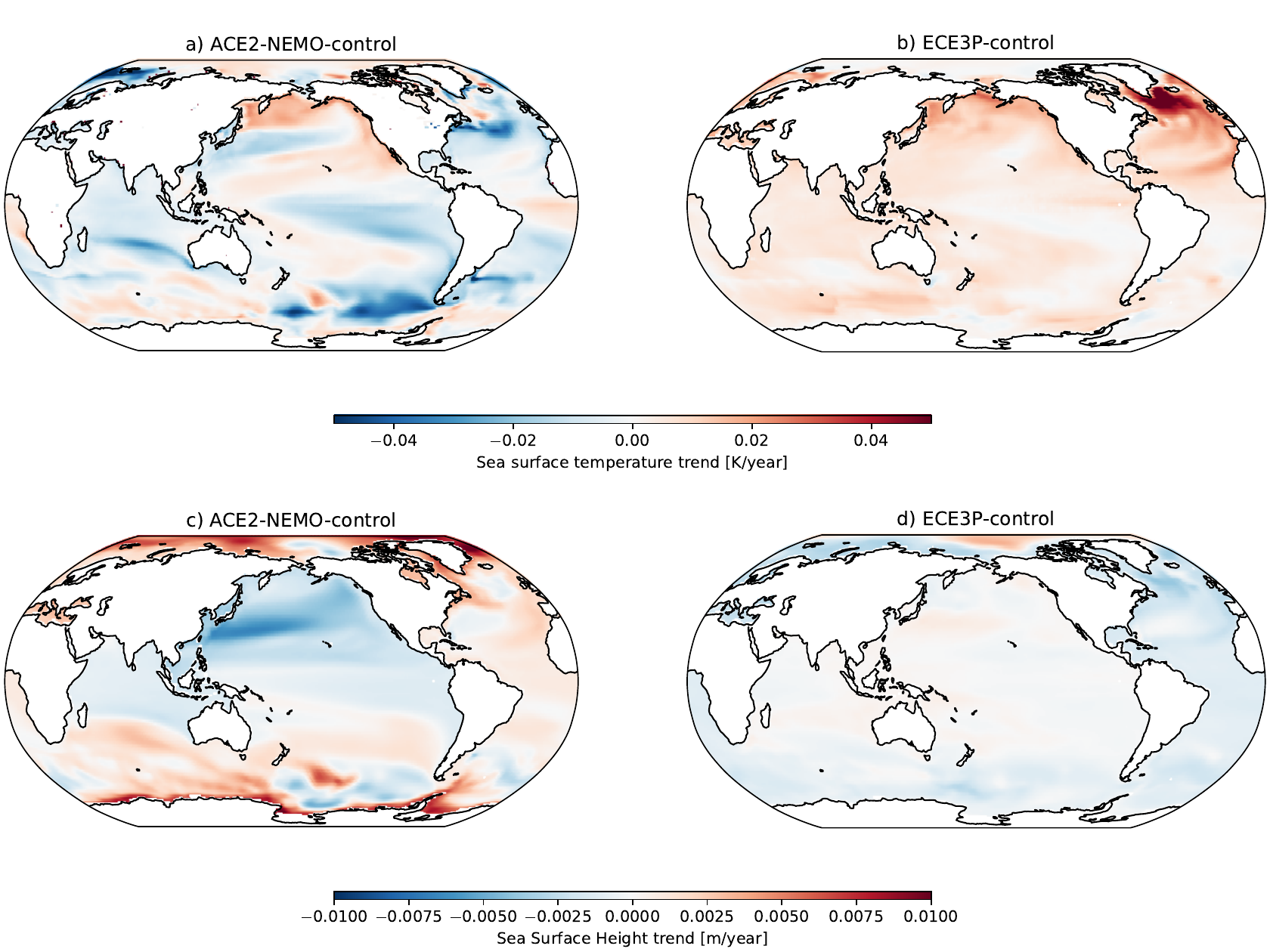}
       \caption{Linear drift in surface variables, estimated by fitting a regression model to each grid cell separately, on yearly aggregated data. Results for the first ensemble member of ACE2-NEMO-control are shown.}
    \label{fig:drift}
\end{figure}

\begin{figure}[!h]
\centering
    \includegraphics[width=0.9\textwidth]{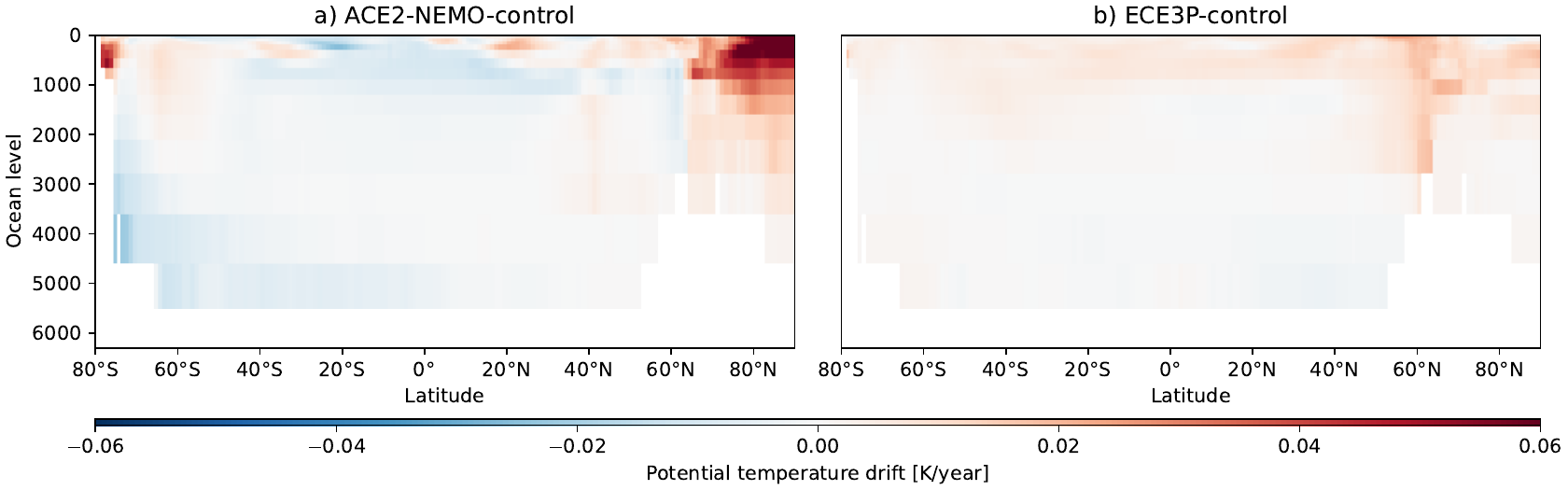}
       \caption{Drift in ocean potential temperature, averaged zonally for each latitude for (a) the first ensemble member of ACE2-NEMO-control and (b) ECE3P-control.}
    \label{fig:drift_toce}
\end{figure}

Maps of changes in sea surface temperature and sea surface height are shown in Fig.~\ref{fig:drift}, to complement the global time series shown in Fig. \ref{fig:global_drift}. These are calculated by fitting a linear regression model for each grid cell separately, for the first ensemble member of ACE2-NEMO-control only. 

The drift in SST (panels a and b) show similar magnitudes between ACE2-NEMO-control and ECE3P-control, but with different patterns; for ACE2-NEMO-control the pattern is similar to the patterns of ENSO heating and cooling \cite{timmermann_nino_2018}, whereas for ECE3P-control the SSTs are consistently warming.

Panels c and d show larger changes in sea surface height for ACE2-NEMO-control compared to ECE3P-control, particularly in the polar regions and in the northern Pacific. The large changes in the polar regions can be explained by the change in sea ice fluxes, which create much larger changes in sea ice melting in the poles (Sec.~\ref{sec:sea_ice}).

Fig.~\ref{fig:drift_toce} shows linear regressions of ocean potential temperature over time for the control runs, where the potential temperatures are first averaged for each latitude. The most significant changes occur at the poles, which is again explained by the changes in sea ice fluxes that create large sea ice variability. Within the rest of the ocean, ACE2-NEMO-control shows a tendency towards cooling, whilst ECE3P-control shows slightly weaker warming tendencies.

\begin{figure}[!t]
\centering
    \includegraphics[width=\textwidth]{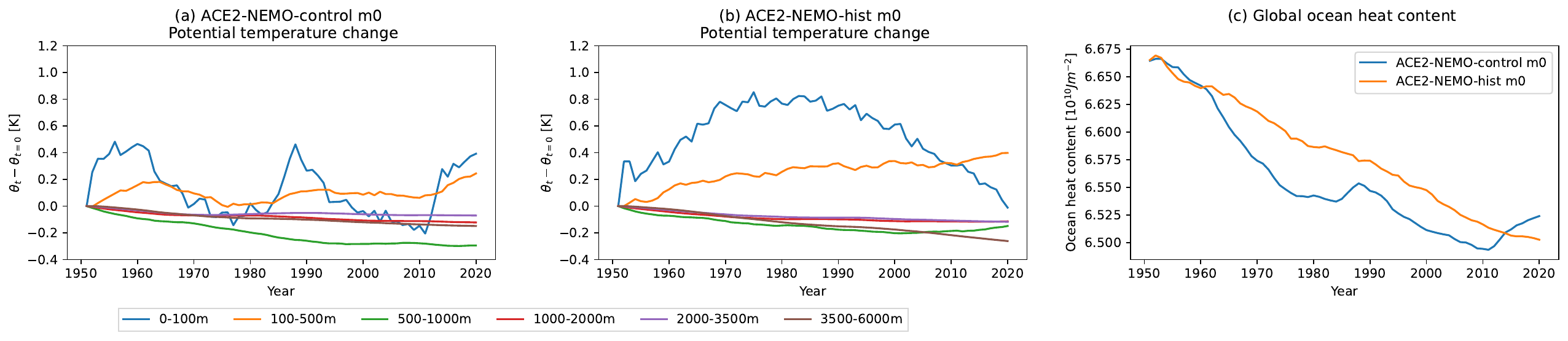}
       \caption{Globally averaged ocean potential temperature $\theta_t$, where temperatures are averaged within depth ranges of 0-100m, 100-500m, 1000-2000m, 2000-3000m, and 3500-6000m (a) ACE2-NEMO-control run (b) ACE2-NEMO-hist run (c) evolution of global ocean heat content for the ACE2-NEMO simulations.}
    \label{fig:potential_temperature_comparison}
\end{figure}

In order to further investigate the model drifts, and the global SST decrease during the ACE2-NEMO-hist run (Fig.~\ref{fig:historical_temp} b), we plot the change in ocean temperature averaged over different depth bands in Fig.~\ref{fig:potential_temperature_comparison}, for ACE2-NEMO-control (panel a) and ACE2-NEMO-hist (panel b). 

For ACE2-NEMO-control (panel a), we see an overall heating trend in ranges 0-100m and 100-500m, modulated by strong decadal variability. The deeper levels show the opposite trend, particularly in the 500-1000m band. The globally averaged heat content for ACE2-NEMO-control (panel c) also shows an overall cooling of the ocean. This points to redistribution of heat from the deeper ocean into the deeper ocean, and an overall cooling, which is most likely due to the ocean model adjusting to a new equilibrium due to the mismatch between NEMO configuration used for the initial conditions and for the ACE2-NEMO runs.

For ACE2-NEMO-hist (panel b), the behaviour in the 0-100m band lacks the decadal variability, and is dominated by the increasing trend until 1980 and decreasing until 2020. The 100-500m band shows a consistent increase throughout, which no changes in trend around 1980 that might suggest significant amounts of heat being transferred to the deeper ocean. There is a slight change in trend in the 500-1000m band around the year 2000, suggesting that there is increased heat being passed to the deeper ocean and so transfer of heat may play some role, but the effect appears to be small relative to the changes in the 0-100m band and occurs much later than the maximum SST seen in Fig.~\ref{fig:historical_temp} (b). Global heat content (panel c) shows a similar overall trend to ACE2-NEMO-control; there is a peak around 1980 that could suggest increased heat uptake into the ocean, however this does not persist for long, and is compensated for by an initially . The change in heat content trend around 2010 perhaps reflects the change in trend seen in the 500-1000m band around the same time.

\section{Sea Ice}
\label{sec:sea_ice}

\begin{figure}[!t]
\centering
    \includegraphics[width=\textwidth]{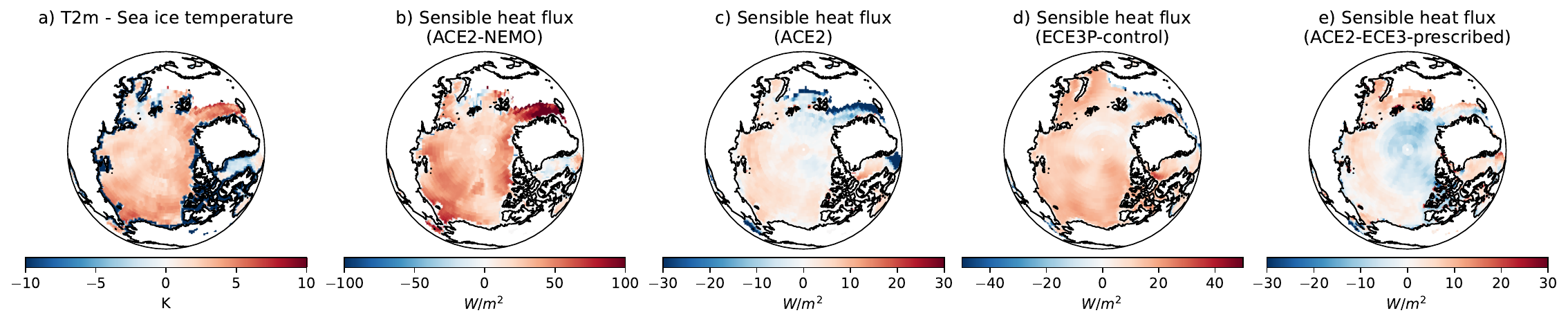}
       \caption{Comparison of sensible heat fluxes over sea ice points, over the first month of simulation only (a) difference between monthly 2-metre temperature and sea surface temperature, (b) monthly averaged sensible heat flux in ACE2-NEMO, using the CORE bulk formulae (c) monthly averaged sensible heat flux from ACE2 outputs (d) sensible heat flux for ECE3P-control (e) monthly averaged sensible heat flux from ACE2-ECE3P-prescribed. Note the different colour bar ranges in plots (b)-(e). }
    \label{fig:shf}
\end{figure}

\begin{figure}[!t]
\centering
    \includegraphics[width=0.9\textwidth]{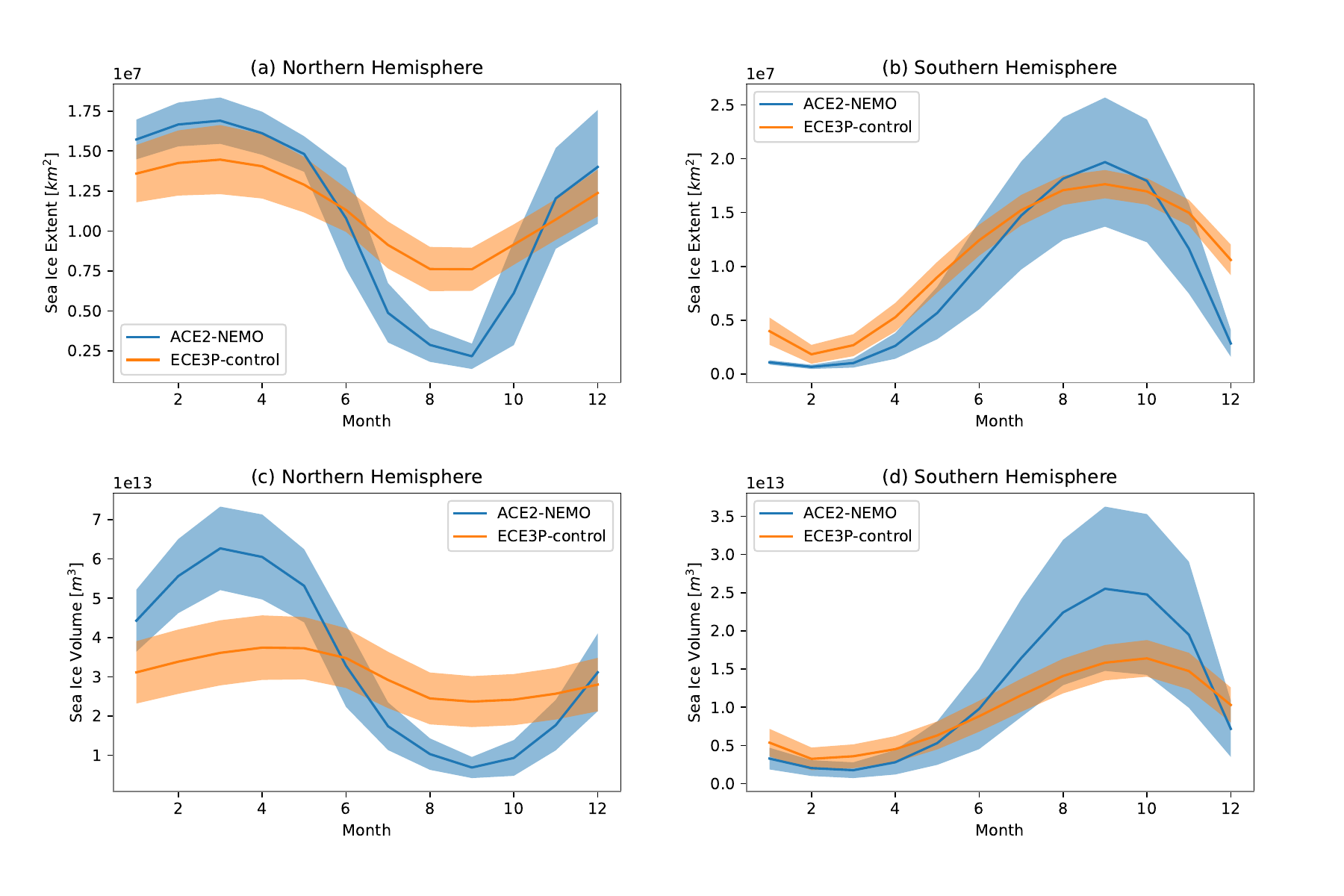}
       \caption{Average annual cycle of sea ice extent (top row) and sea ice volume (bottom row) for the northern hemisphere (left column) and southern hemisphere (right column). }
    \label{fig:seasonal_cycle_ice}
\end{figure}

Fig.~\ref{fig:shf} compares the surface heat gradients over ice to the resultant sensible heat flux. Panel a shows the temperature gradient between 2-metre temperature and bulk sea surface temperature averaged over the first month of simulation, whilst panels b-e show the sensible heat fluxes for different models. Panel b shows the sensible heat fluxes when using the CORE bulk formulae in ACE2-NEMO-control, showing an agreement between the sign of sensible heat fluxes and temperature gradient. Panel c shows the same for the flux output by ACE2, which shows inconsistencies particularly to the East of Greenland; inconsistencies also persist when using prescribed sea surface conditions (panel e). Panel d shows that ECE3P-control shows similar behaviour to ACE2-NEMO-control.

The annual cycle of sea ice volume for the northern and southern hemisphere is shown in Fig.~\ref{fig:seasonal_cycle_ice} for ACE2-NEMO-control and ECE3P-control. The maxima and minima of the cycle for ACE2-NEMO-control align well with ECE3P-control, indicating the ice fluxes are correctly modulating the sea ice, however it is clear that the variation in sea ice volume and extent is much larger for ACE2-NEMO. The variation in sea ice volume in Fig.~\ref{fig:seasonal_cycle_ice} (c) and (d) indicates that the sea ice becomes much thicker in winter and much thinner in summer. This difference may be explained by the use of bulk sea ice temperature in the calculation of sensible fluxes over ice for ACE2-NEMO, since this is the field output from NEMO. This may give larger temperature gradients than using surface ice temperature. Plots comparing sensible heat flux in Fig.~\ref{fig:shf} also confirm that this heat flux is much larger than the original ACE2 fluxes or the ECE3P-control experiment. 

\begin{figure}[!t]
\centering
    \includegraphics[width=0.9\textwidth]{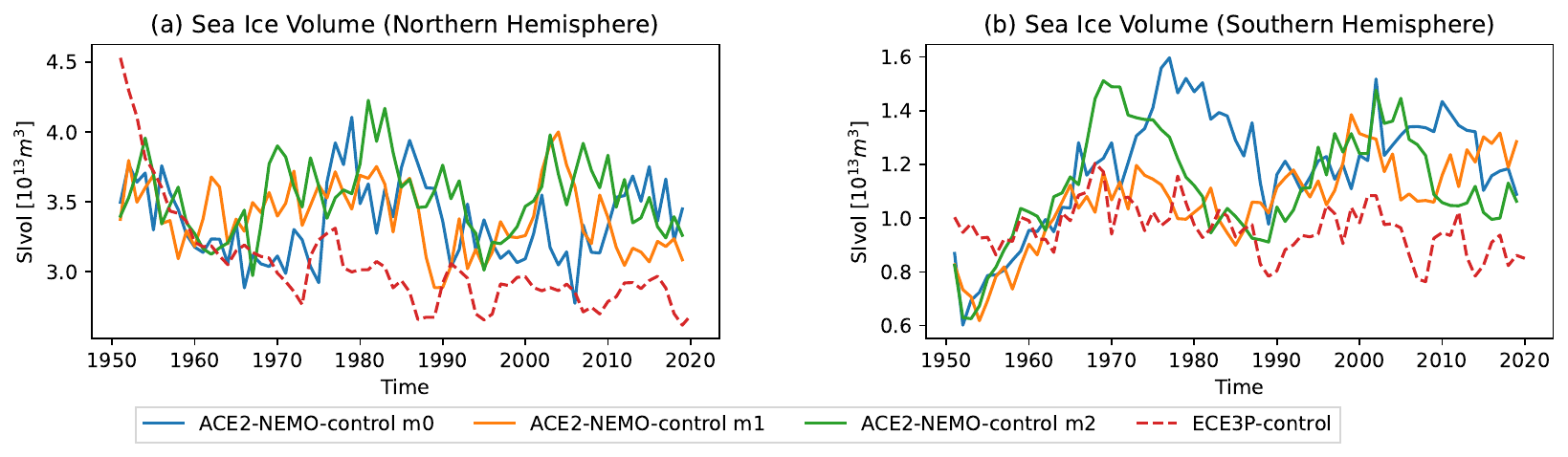}
       \caption{Sea ice volume time series for the ACE2-NEMO-control 3-member ensemble and EC-Earth3P-control averaged over (a) northern hemisphere, (b) southern hemisphere.}
    \label{fig:sea_ice_evolution}
\end{figure}

Sea ice changes broken down by hemisphere are shown in Fig.~\ref{fig:sea_ice_evolution}, providing a breakdown of the global sea ice volume shown in Fig.~\ref{fig:global_drift} (b). From this we can see that the decrease in sea ice volume for ECE3P-control is driven by behaviour in the Arctic (panel a). Additionally, there is an initial increase in sea ice for ACE2-NEMO-control seen around the Antarctic for the first decade (panel b).

\begin{figure}
\centering
    \includegraphics[width=0.5\textwidth]{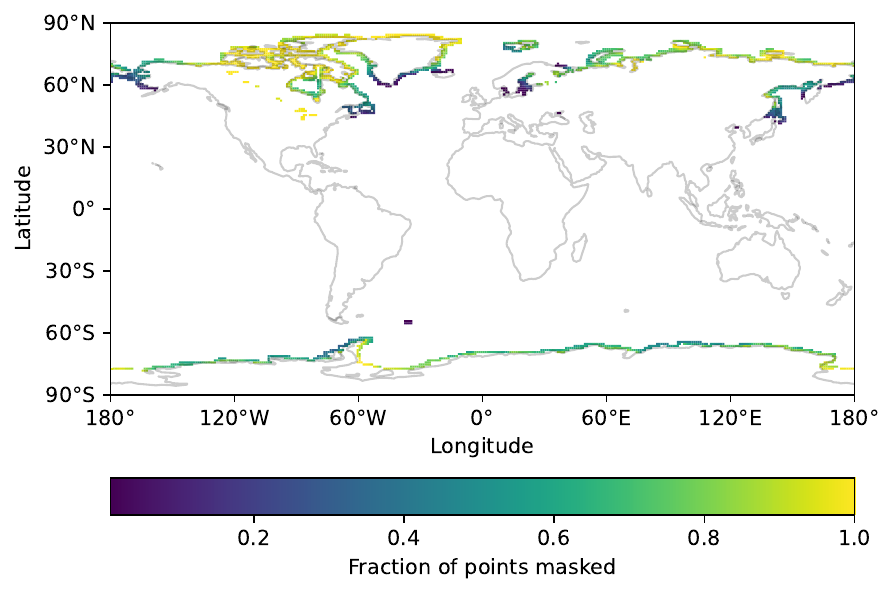}
       \caption{Map showing the fraction of times that particular points are masked by the coastal ice flux correction. }
    \label{fig:masked_fraction}
\end{figure}

To illustrate the spatial extent of the coastal ice flux masking that stabilises the model, a map showing the fraction of masked points is shown in Fig.~\ref{fig:masked_fraction}. For each grid square, this is calculated as the fraction of months for which that grid square would have had the masking applied. The area fraction of grid points over the ocean that have a masking more than 10\% of the time is approximately 3\%. 

\section{El Ni\~{n}o Southern Oscillation Analysis}
\label{sec:enso_supp}

\begin{figure}[!h]
\centering
    \includegraphics[width=0.9\textwidth]{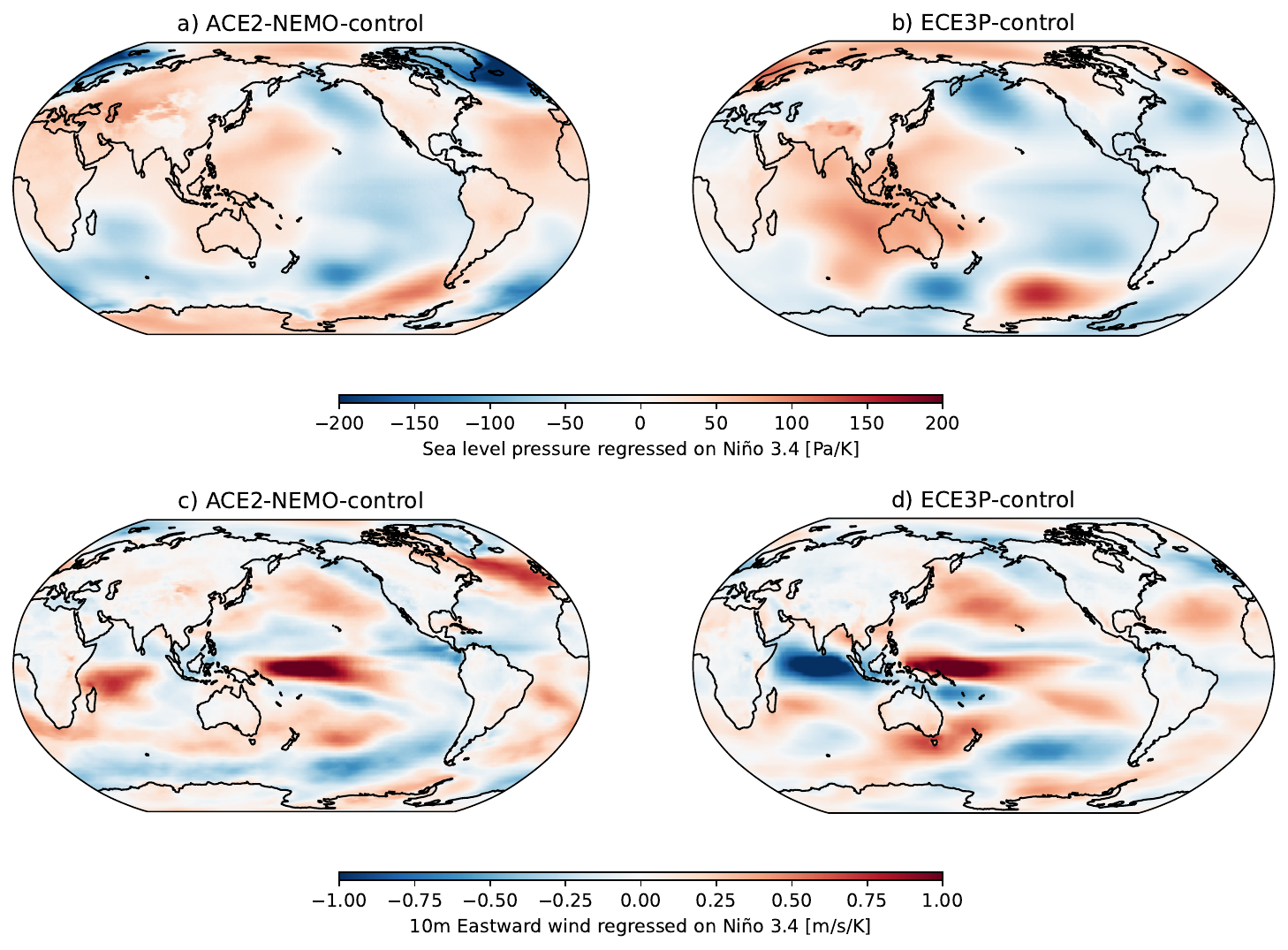}
       \caption{Regression of monthly surface variables onto the monthly Ni\~{n}o 3.4 index (a) and (b): sea level pressure, (c) and (d) 10-metre eastward wind component. }
    \label{fig:enso_mslp_10mU}
\end{figure}

In this subsection we provide additional analysis of the coupled El Ni\~{n}o Southern Oscillation (ENSO) variability in ACE2-NEMO. A regression of the Ni\~{n}o 3.4 index against surface variables is shown in Fig.~\ref{fig:enso_mslp_10mU}, for sea level pressure (panels a and b) and 10-metre eastward wind component (panels c and d). Whilst the sea-level pressure response of ACE2-NEMO-control (panel a) is similar to ECE3P-control (panel b), there are clear differences over the Indian and Southern Oceans, and northern Atlantic, and an overall lower magnitude in the strength of the response. ACE2-NEMO-control captures some of the 10-metre winds response to Ni\~{n}o 3.4 in the tropical Pacific (panel c), although the response is shifted eastward compared to ECE3P-control (panel d). There is also a large discrepancy over the Indian Ocean, where ECE3P-control shows a strong westward wind response, whereas ACE2-NEMO-control shows a weak eastward wind response.

\section{ Verification of flux calculations}

In this section we explore the quality of the flux calculation, in order to highlight any biases that may contribute to the behaviour of the coupled model. We first approach this by running the coupling calculation using ERA5 atmospheric and oceanic variables as input; this `perfect model' setup ensures that we can assess the quality of the coupling in the absence of any ACE2 errors.

Simulations are initialised on the first of every month, for the years 1951-1955, at hours 00:00, 06:00, 12:00 and 18:00. Each simulation is run to a lead time of 6 hours, to ensure that comparison with ERA5 data is still meaningful. Areas with a sea ice fraction of more than 10\% are masked out, since the AirSeaFluxCode calculation is only applicable to ocean points. Fluxes follow the NEMO sign convention that positive fluxes indicate a gain of that quantity for the ocean.

\begin{figure}[htbp!]
\centering
    \includegraphics[width=0.9\textwidth]{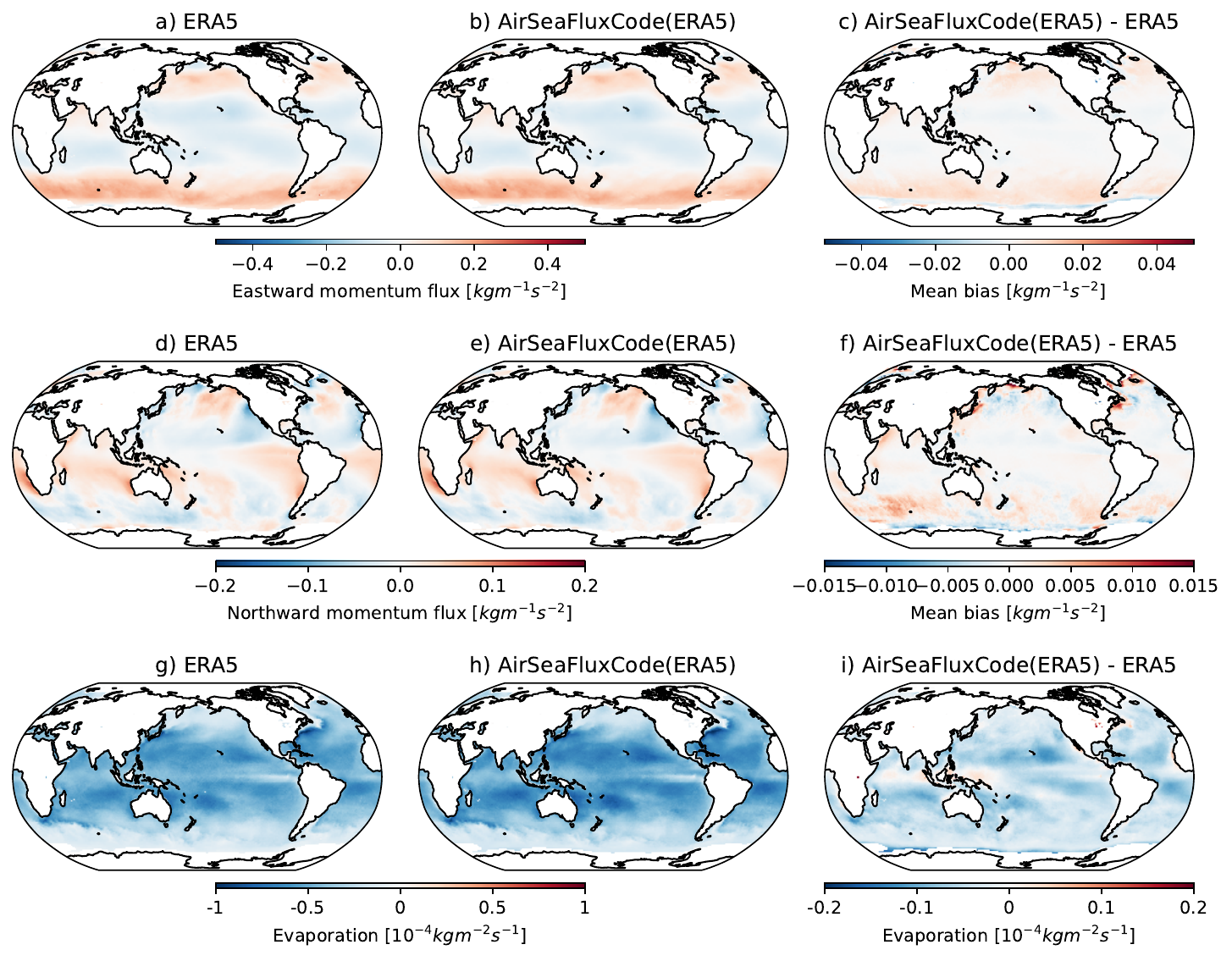}
       \caption{Comparison of the momentum fluxes and evaporation for ERA5 (left column), and for the AirSeaFluxCode calculation using ERA5 state variables as input (middle column, labelled as AirSeaFluxCode(ERA5)). Mean bias is shown in the right column. }
    \label{fig:momentum_flux_comparison}
\end{figure}

We first investigate the quality of the momentum fluxes: Fig.~\ref{fig:momentum_flux_comparison} compares eastward and northward wind stress (panels a and d) and evaporation (panel g) with the same fluxes calculated using AirSeaFluxCode with ERA5 variables as input (panels b, e, and h). Differences between the fluxes are shown in panels c, f, and i (note the much smaller colorbar ranges used for these bias plots). Overall, we can see good agreement between the AirSeaFluxCode and ERA5 momentum fluxes. The most significant biases in eastward momentum flux can be seen in the midlatitudes, concentrated in areas where the momentum flux is largest. The biases in northward momentum flux are generally smaller in magnitude than the Eastward flux, and show highest values particularly near coastal points such as Japan, the northern Pacific, and the North-West Atlantic. Whilst some differences remain to be improved on, this analysis confirms that the momentum flux calculation is very consistent with the ERA5 method of calculating fluxes, and by extension means that it is reasonably consistent with the flux calculation of ACE2, which is trained to approximate ERA5. For evaporation, we observe that the AirSeaFluxCode calculation produces a larger evaporation than ERA5, with biases of magnitude up to around 20\%.

\begin{figure}[htbp!]
\centering
    \includegraphics[width=0.9\textwidth]{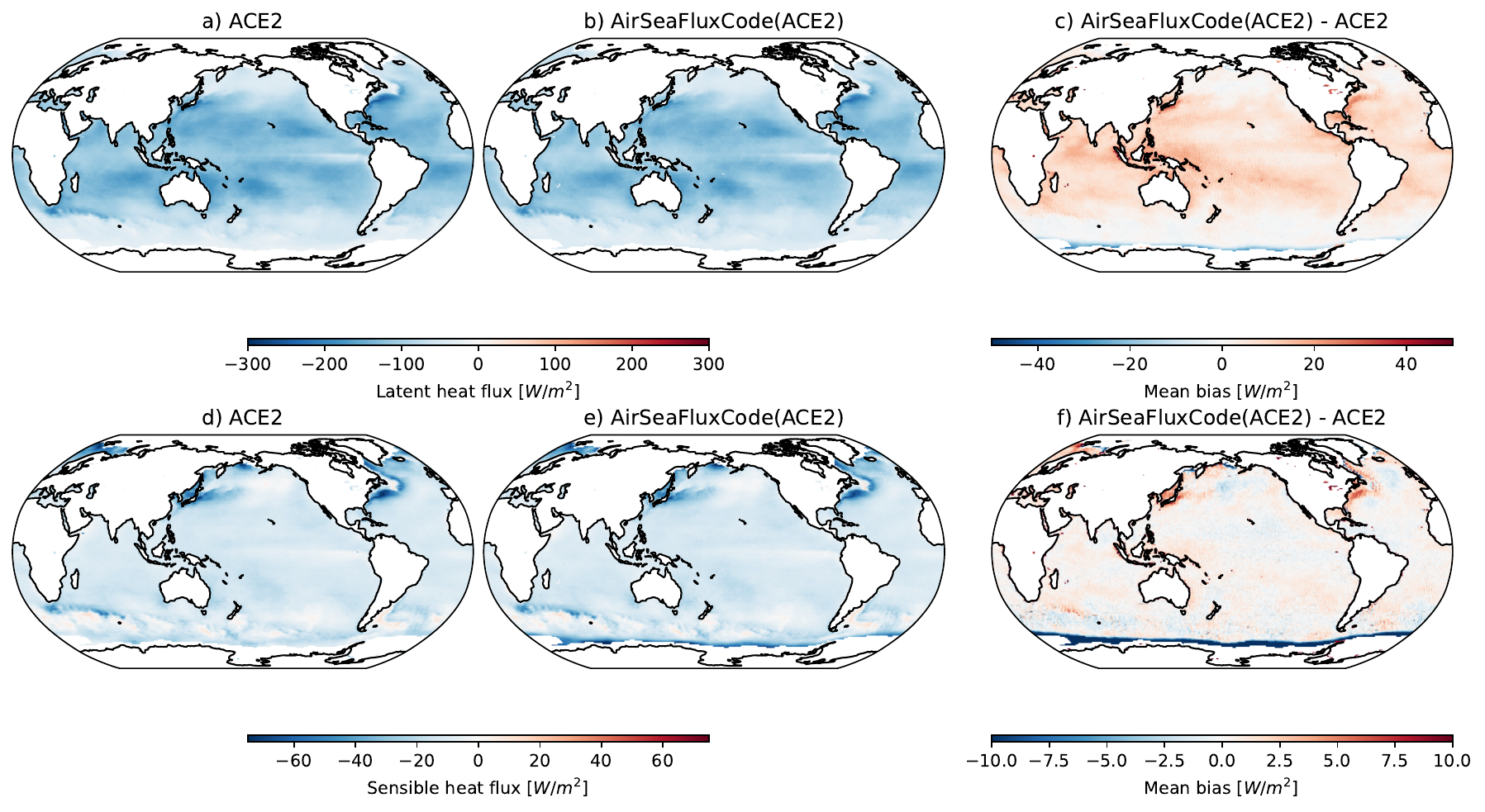}
       \caption{Comparison of the heat fluxes ACE2 (left column), and for the AirSeaFluxCode calculation using ACE2 state variables as input (middle column, labelled as AirSeaFluxCode(ACE2)). Mean bias is shown in the right column. }
    \label{fig:flux_consistency_comparison}
\end{figure}

We probe this further, by comparing the fluxes produced by ACE2, with those produced by AirSeaFluxCode using ACE2 variables as input. This comparison allows an assessment of how consistent the flux calculation of AirSeaFluxCode is with the implicit flux calculation of ACE2. Results are shown in Fig.~\ref{fig:flux_consistency_comparison}, with mean biases shown in the right hand column (note the much smaller colorbar ranges used for these bias plots). Overall we see very good consistency between the ACE2 fluxes and the AirSeaFluxCode fluxes, indicating a strong similarity between two calculations. Note the large differences seen at the ice edge for panels c and f: this is due to AirSeaFluxCode calculating fluxes based on sea surface temperature, which for ERA5 is set to the freezing point of salt water beneath sea ice points, so this is an artefact of sea ice temperature not being incorporated into the AirSeaFluxCode calculation rather than a bias.

\begin{figure}[htbp!]
\centering
    \includegraphics[width=0.9\textwidth]{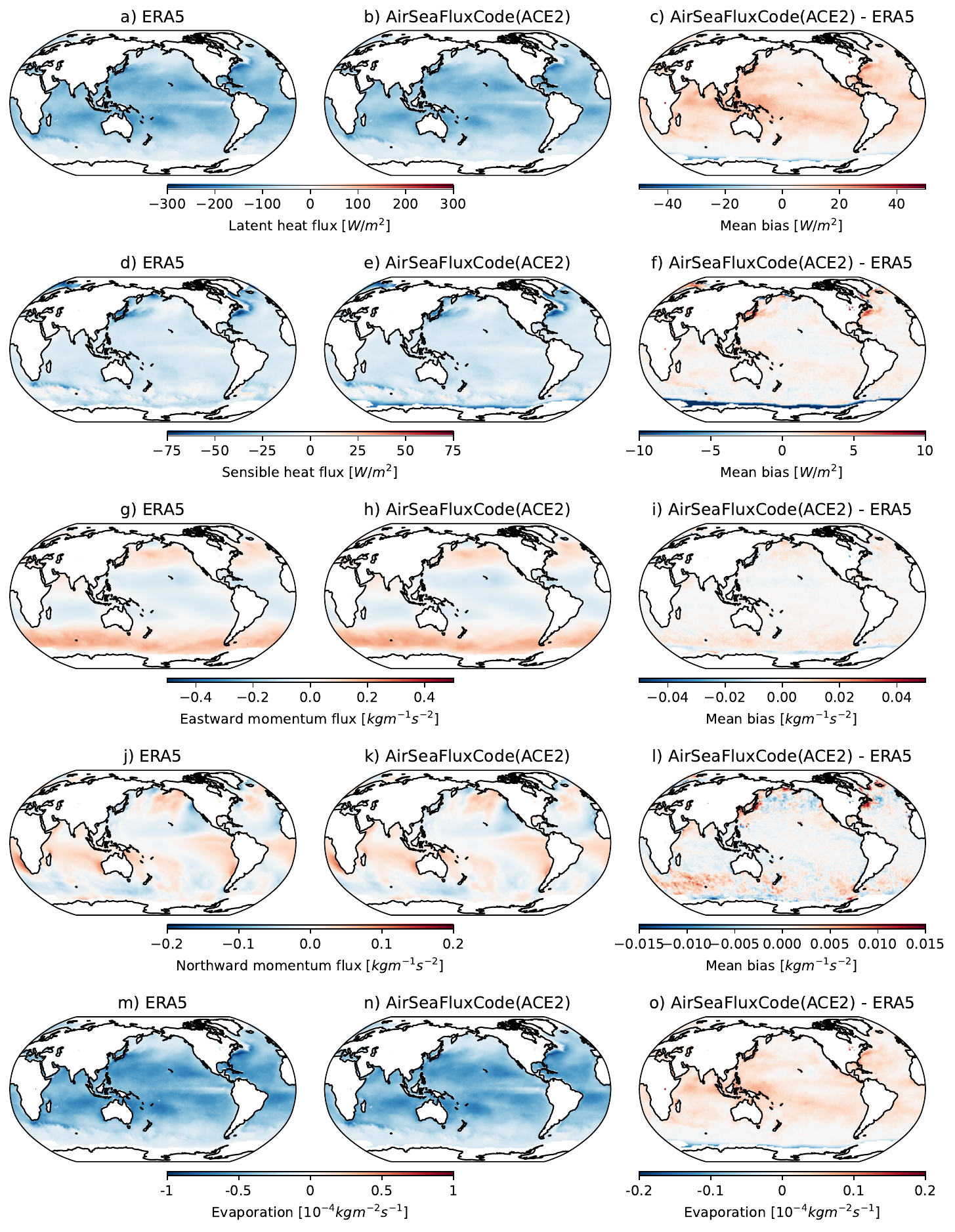}
       \caption{Comparison of the heat fluxes, momentum fluxes, and evaporation for ERA5 (left column), and for the AirSeaFluxCode calculation using ACE2 state variables as input (middle column, labelled as AirSeaFluxCode(ACE2)). Mean bias is shown in the right column. }
    \label{fig:momentum_flux_comparison_ace2}
\end{figure}

Our final analysis of these fluxes compares ERA5 fluxes with fluxes calculated using AirSeaFluxCode with ACE2 variables as input. This comparison allows us to assess total biases when combining the effects of AirSeaFluxCode and ACE2. Results are shown in Fig.~\ref{fig:momentum_flux_comparison_ace2}. The overall magnitude of the differences for the heat fluxes (panel c and f) and momentum fluxes (panels i and l) are very similar to those observed in Fig.~\ref{fig:flux_consistency_comparison} and Fig.~\ref{fig:momentum_flux_comparison}; this suggests that the ACE2 inputs are well aligned with ERA5 and do not add a significant extra bias compared to the flux calculation. A notable exception to this is the evaporation (panels m-o), which show a net positive bias (i.e. corresponding to net reduction in the evaporation magnitude of evaporation, since fluxes are positive downward) in contrast to the negative bias seen in Fig.~\ref{fig:momentum_flux_comparison}. This bias is in the wrong direction to be able to explain the hydrological biases seen in Sec.~\ref{sec:sensitivity}, where a negative drift is observed in sea surface height.

\begin{figure}[htbp!]
\centering
    \includegraphics[width=0.9\textwidth]{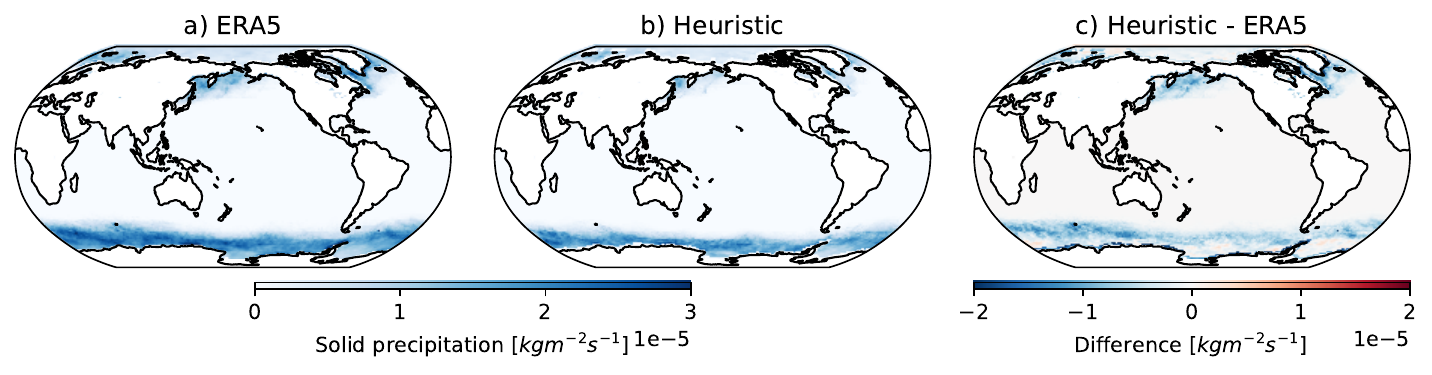}
       \caption{Comparison of solid precipitation from ERA5 (panel a), using the heuristic method (panel b). Mean bias is shown in panel c. }
    \label{fig:solid_precip_comparison}
\end{figure}

An assessment of the bias in solid precipitation is shown in Fig.~\ref{fig:solid_precip_comparison}. From this we can see that the spatial pattern of solid precipitation produced by the heuristic calculation is too concentrated in many areas, and does not extend enough into the Southern Ocean, leading to positive biases near the coast of Antarctica and over the Arctic Ocean, and negative biases at lower latitudes. 

\end{document}